\documentclass[pdflatex,sn-mathphys-num]{sn-jnl}

\usepackage{graphicx}%
\usepackage{multirow}%
\usepackage{amsmath,amssymb,amsfonts}%
\usepackage{amsthm}%
\usepackage{mathrsfs}%
\usepackage[title]{appendix}%
\usepackage{xcolor}%
\usepackage{textcomp}%
\usepackage{manyfoot}%
\usepackage{booktabs}%
\usepackage{algorithm}%
\usepackage{algorithmicx}%
\usepackage{algpseudocode}%
\usepackage{listings}%
\usepackage{bbold}
\usepackage{slashed}
\usepackage{subfigure}
\usepackage{braket,bm}
\usepackage{cancel}
\usepackage{dsfont}
\usepackage{tikz}
\usepackage{array}
\usepackage{overpic}
\usepackage{float}
\usepackage{threeparttable}
\usepackage{orcidlink}

\graphicspath{{figs/}}

\begin{document}

\title{Nucleon Structure from Basis Light-Front Quantization : Status and Prospects}

\author[1]{\fnm{James P.} \sur{Vary} \orcidlink{0000-0002-3500-4314}
}\email{jvary@iastate.edu}

\author[2,3,4,*]{\fnm{Chandan} \sur{Mondal} \orcidlink{0000-0002-0000-5317}
}\email{mondal@impcas.ac.cn}

\author[1,2,3]{\fnm{Siqi} \sur{Xu} \orcidlink{0000-0001-5570-6340}
}\email{xsq234@impcas.ac.cn}

\author[2,3,4,5]{\fnm{Xingbo} \sur{Zhao} \orcidlink{0000-0001-5568-054X}
}\email{xbzhao@impcas.ac.cn}

\author[6,7]{\fnm{Yang} \sur{Li} \orcidlink{0000-0001-9446-6503}
}\email{leeyoung1987@ustc.edu.cn}

\author[ ]{\fnm{(BLFQ Collaboration)}}

\affil[1]{\orgdiv{Department of Physics and Astronomy}, \orgname{Iowa State University}, \orgaddress{\city{Ames}, \state{Iowa}, \postcode{IA 50011}, \country{USA}}}

\affil[2]{\orgdiv{Institute of Modern Physics}, \orgname{Chinese Academy of Sciences}, \orgaddress{\city{Lanzhou}, \state{Gansu}, \postcode{730000}, \country{China}}}

\affil[3]{\orgdiv{School of Nuclear Physics}, \orgname{University of Chinese Academy of Sciences}, \orgaddress{\city{Beijing}, \postcode{100049}, \country{China}}}

\affil[4]{\orgdiv{CAS Key Laboratory of High Precision Nuclear Spectroscopy, Institute of Modern Physics}, \orgname{Chinese Academy of Sciences}, \orgaddress{\city{Lanzhou}, \state{Gansu},\postcode{730000}, \country{China}}}

\affil[5]{\orgdiv{Advanced Energy Science and Technology}, \orgname{Guangdong Laboratory}, \orgaddress{\city{Guangdong}, \postcode{516000}, \country{China}}}

\affil[6]{\orgname{University of Science and Technology of China}, \orgaddress{\city{Hefei}, \state{Anhui}, \postcode{230026}, \country{China}}}

\affil[7]{\orgdiv{Anhui Center for Fundamental Sciences in Theoretical Physics}, \orgname{University of Science and Technology of China}, \orgaddress{\city{Hefei}, \postcode{230026}, \country{China}}}

\abstract{We review recent advancements in understanding nucleon structure within the Basis Light-Front Quantization (BLFQ) framework—a fully relativistic, nonperturbative approach to solving quantum field theories. In its initial phase, we start with the leading Fock sector $|qqq\rangle$ and an effective light-front Hamiltonian incorporating confinement and one-gluon exchange within which BLFQ  can already successfully describe key nucleon observables. The framework has since been extended to include the next-to-leading Fock sector $|qqqg\rangle$, enabling studies of gluonic contributions to the nucleon's internal structure, including gluon helicity, orbital angular momentum, and three-dimensional imaging through generalized and transverse momentum dependent parton distributions (GPDs and TMDs). Most recently, BLFQ has achieved a significant milestone by computing nucleon light-front wavefunctions as eigenstates of the QCD Hamiltonian without an explicit confining potential. These calculations, including Fock sectors up to $|qqqq\bar{q}\rangle$, further develop the path to first-principles predictions of quark and gluon matter densities, helicity and transversity distributions, and spin observables, showing qualitative agreement with experimental and phenomenological results. Together, these developments highlight BLFQ’s growing capacity to provide an increasingly complete and realistic picture of nucleon structure grounded in QCD.}

\keywords{Nucleon, Light-front quantization, Form factors, PDFs, GPDs, Spin decomposition}

\maketitle

\tableofcontents

\clearpage

\section{Introduction}
A central goal in hadron physics is to understand how the properties of nucleons and other hadrons emerge from their quark and gluon constituents. This question has driven extensive experimental and theoretical efforts over several decades such as high-energy lepton scattering experiments. While the main focus of this review is the recent progress in developing and applying a relativistic theoretical framework based on light-front quantization for the proton's structure, we first overview recent key experimental and theoretical results available for comparison.

Electromagnetic form factors (FFs), accessible via elastic scattering, encode information about the spatial distribution of charge and magnetization within the nucleon through their Fourier transforms.
Deep inelastic lepton scattering (DIS), on the other hand, provides access to parton distribution functions (PDFs), which describe the longitudinal momentum distributions of quarks and gluons. While FFs and PDFs have significantly advanced our understanding, they offer only limited insight into the nucleon’s full three-dimensional structure.
Generalized parton distributions (GPDs), which appear in hard exclusive processes such as deeply virtual Compton scattering (DVCS) and vector meson production, bridge this gap. GPDs provide a unified description that combines spatial and momentum information, revealing the distribution and orbital motion of partons.
Complementarily,  the transverse-momentum-dependent parton distribution functions (TMDs) provide information on the distribution of the transverse momentum in addition to the longitudinal momentum fraction. The TMDs, especially of quarks, can be extracted from the cross section associated with a specific azimuthal angle in the semi-inclusive deep inelastic scattering (SIDIS) processes or the Drell-Yan processes. These two classes of distribution functions allow us to draw a  three-dimensional imaging of the nucleon.

The matrix element of the electromagnetic current for the nucleon is parameterized by two independent form factors: the Dirac and Pauli form factors. These nucleon FFs have been the focus of extensive experimental and theoretical investigations over the years~\cite{Perdrisat:2006hj,Pacetti:2014jai,Punjabi:2015bba}. Experimentally, two well-established methods are used to extract FFs. The first is the Rosenbluth separation technique, which analyzes unpolarized scattering cross sections to obtain the Sachs form factors—linear combinations of the Dirac and Pauli FFs. The second is the polarization transfer method, which uses either a polarized target or a recoiling polarized proton in combination with a polarized lepton beam to directly measure the ratio of Sachs FFs.
However, inconsistencies have been observed between these two methods, particularly in the extraction of the ratio of the proton’s electric to magnetic Sachs FFs. In double-polarization experiments~\cite{JeffersonLabHallA:2000dxx,Punjabi:2005wq,JeffersonLabHallA:2001qqe,Puckett:2011xg,Puckett:2010ac}, this ratio decreases almost linearly with increasing momentum transfer ($Q^2 > 0.5~\text{GeV}^2$), while results from Rosenbluth separation~\cite{Hand:1963zz,Janssens:1965kd,Price:1971zk,Litt:1969my,Berger:1971kr,Bartel:1973rf,Borkowski:1974mb,Simon:1980hu,Walker:1993vj,Andivahis:1994rq,E94110:2004lsx,Qattan:2004ht} remain nearly constant in the spacelike region.
This apparent discrepancy has been a long-standing puzzle, which may be largely resolved by accounting for the contribution of two-photon exchange (TPE) processes that are neglected in the one-photon exchange approximation of the Rosenbluth method~\cite{Klest:2025yik}. In particular, TPE effects modify the unpolarized cross section more significantly than polarization observables, thereby reconciling the two extraction techniques when higher-order corrections are consistently included.
For the neutron, experimental data at large momentum transfers ($Q^2 = -q^2$) are scarce, and existing theoretical models often struggle to describe its form factors. Fortunately, high-precision measurements of both proton and neutron FFs are expected from ongoing and future experiments at Jefferson Lab. Theoretically, nucleon electromagnetic FFs have been studied using various frameworks~\cite{Alexandrou:2017ypw,QCDSF:2017ssq,Alexandrou:2018sjm,Shintani:2018ozy,He:2017viu,Alarcon:2017lhg,Abidin:2009hr,Mondal:2016xpk,Sufian:2016hwn,Brodsky:2014yha,Mondal:2015uha,Chakrabarti:2013dda,Ye:2017gyb,Gutsche:2013zia,Cloet:2012cy,Pasquini:2007iz,Ahmady:2021qed,Cheng:2024cxk,Mondal:2019jdg,Xu:2021wwj,Xu:2024sjt}, and the flavor decomposition of FFs has been addressed in Refs.~\cite{Cates:2011pz,Qattan:2012zf,Diehl:2013xca}.

In contrast to the nucleon electromagnetic FFs, which have been extensively studied experimentally, our knowledge of the axial FFs remains limited. To date, only two types of experiments provide access to axial FFs: (anti)neutrino scattering off protons or nuclei, and charged pion electroproduction. A comprehensive review of the available experimental data on nucleon axial FFs can be found in Refs.~\cite{Bernard:2001rs,Schindler:2006jq}.
Axial FFs have also been investigated using various theoretical approaches~\cite{Tsushima:1988xv,Julia-Diaz:2004yyg,Mamedov:2016ype,Liu:2016kpb,Anikin:2016teg,Adamuscin:2007fk,Aznauryan:2012ba,Ramalho:2017tga,Hashamipour:2019pgy,Mondal:2017wbf}. In recent years, lattice QCD simulations of axial FFs have been performed for pion masses in the range $0.2–0.6$ GeV~\cite{Alexandrou:2013joa,Liang:2016fgy,Green:2017keo,Yao:2017fym,Abdel-Rehim:2015owa,Bali:2014nma,Bhattacharya:2016zcn}. Notably, a recent study by the PACS Collaboration~\cite{Ishikawa:2018rew} reports a lattice QCD calculation of the nucleon axial FF using $2+1$ flavors near the physical pion mass.

PDFs encode the nonperturbative structure of the nucleon by describing the number densities of its confined constituents---quarks and gluons---as functions of the light-front longitudinal momentum fraction $x$ carried by each constituent. At leading twist, the nucleon’s full spin structure is characterized by three independent PDFs: the unpolarized distribution $f_1(x)$, the helicity distribution $g_1(x)$, and the transversity distribution $h_1(x)$.
A precise determination of PDFs is essential for interpreting high-energy scattering experiments in the LHC era. Global analysis efforts by collaborations such as NNPDF~\cite{NNPDF:2017mvq}, HERAPDF~\cite{Alekhin:2017kpj}, MMHT~\cite{Harland-Lang:2014zoa}, CTEQ~\cite{Dulat:2015mca}, and MSTW~\cite{Martin:2009iq} have significantly advanced our knowledge of PDFs and their uncertainties. The helicity distributions, in particular, have been extracted with increasing precision from polarized lepton-proton inclusive scattering~\cite{Aidala:2012mv,Deur:2018roz}.
Perturbative QCD (pQCD) predicts that the ratio of polarized to unpolarized PDFs for the up quark approaches unity as $x \to 1$~\cite{Brodsky:1994kg,Avakian:2007xa}. For the down quark, this ratio remains negative across the experimentally accessible range $x \lesssim 0.6$~\cite{JeffersonLabHallA:2003joy,JeffersonLabHallA:2004tea,JeffersonLabHallA:2014mam,CLAS:2006ozz,HERMES:2003gbu,HERMES:2004zsh,COMPASS:2010hwr}, with no observed sign reversal. This trend is supported by global pQCD analyses~\cite{deFlorian:2008mr,deFlorian:2009vb,Nocera:2014gqa,Jimenez-Delgado:2014xza,Ethier:2017zbq}, Dyson-Schwinger equations (DSEs)~\cite{Roberts:2013mja}, and light-front (LF) quantization~\cite{Xu:2022yxb} studies. However, a recent prediction based on light-front holography suggests a possible sign reversal for the polarized down-quark distribution at large $x$~\cite{Liu:2019vsn}.
Lattice QCD has also been employed to study nucleon PDFs using a variety of methods, including the path-integral formulation~\cite{Liu:1999ak,Liu:1993cv}, inversion techniques~\cite{QCDSF:2012mkm,Chambers:2017dov}, pseudo-PDFs~\cite{Radyushkin:2017cyf,Orginos:2017kos}, quasi-PDFs~\cite{Ji:2013dva,Lin:2014zya,Alexandrou:2015rja,Alexandrou:2016jqi}, and lattice cross-section approaches~\cite{Ma:2017pxb}. A detailed overview of the current progress and remaining challenges in lattice-based extractions of PDFs is provided in Ref.~\cite{Lin:2017snn}.

While the unpolarized and helicity PDFs have been extensively studied and are relatively well determined, the transversity PDF, $h_1(x)$, remains less understood. Its chiral-odd nature prevents it from contributing to inclusive deep inelastic scattering; instead, it can only be accessed through processes involving another chiral-odd partner~\cite{Jaffe:1991kp}.
The $h_1$ distribution encodes the correlation between the transverse polarization of quarks and that of the nucleon. A primary method to probe $h_1$ is through Collins azimuthal asymmetries in semi-inclusive deep inelastic scattering (SIDIS)~\cite{Collins:1992kk}. Significant experimental efforts have been devoted to measuring these asymmetries, including by the HERMES~\cite{HERMES:2004mhh,HERMES:2010mmo}, JLab Hall A~\cite{JeffersonLabHallA:2011ayy}, and COMPASS~\cite{COMPASS:2012ozz} Collaborations. Complementary measurements of azimuthal asymmetries in back-to-back hadron production from $e^+e^-$ annihilation have been performed by the BELLE and BABAR Collaborations~\cite{Belle:2005dmx,Belle:2008fdv}.
Extracting the transversity distribution requires knowledge of the chiral-odd Collins fragmentation functions~\cite{Collins:1992kk,Boer:2003ya,Bacchetta:2008wb,Courtoy:2012ry}. In recent years, global analyses have been performed using SIDIS and $e^+e^-$ data to extract transversity~\cite{Anselmino:2013vqa,Kang:2015msa,Radici:2015mwa,Bacchetta:2011ip,Bacchetta:2012ty}. A comprehensive extraction based on both electron-proton and proton-proton scattering data has been reported in Ref.~\cite{Radici:2018iag}. The first extraction of transversity PDFs through a QCD global analysis of $e^+e^-$ annihilation, SIDIS, and $pp$ dihadron measurements has been recently reported by the JAM Collaboration~\cite{Cocuzza:2023oam}.

GPDs provide crucial insights into the spin and orbital angular momentum of the nucleon’s constituents, as well as its spatial structure. Unlike ordinary PDFs, GPDs depend on three variables: the longitudinal momentum fraction $x$, the momentum transfer squared $t$, and the skewness parameter $\xi$, which represents the longitudinal momentum transfer. For comprehensive reviews, see Refs.~\cite{Diehl:2003ny,Belitsky:2005qn,Goeke:2001tz,Ji:1998pc,Burkardt:2002hr}.

In the forward limit ($t=0$), GPDs reduce to standard PDFs. Their first moments correspond to electromagnetic form factors, while the second moments—through Ji’s sum rule—are connected to the total angular momentum carried by quarks and gluons~\cite{Ji:1996ek}. Although GPDs are off-forward matrix elements and lack a probabilistic interpretation, their Fourier transforms with respect to purely transverse momentum transfer ($\xi=0$) yield impact parameter dependent GPDs. These distributions admit a probabilistic interpretation, satisfy positivity constraints~\cite{Ralston:2001xs}, and provide spatial imaging of partons in the transverse plane for fixed longitudinal momentum fraction $x$.


A precise understanding of GPDs is essential for the analysis and interpretation of various exclusive scattering processes. These include DVCS~\cite{Ji:1996nm,Goeke:2001tz}, deeply virtual meson production (DVMP)~\cite{Goloskokov:2007nt,Collins:1996fb,Goloskokov:2024egn}, neutrino and electroweak meson production~\cite{Pire:2017yge,Pire:2021dad}, single diffractive hard exclusive processes (SDHEPs)~\cite{Qiu:2022pla,Grocholski:2022rqj,Duplancic:2022ffo}, timelike Compton scattering (TCS)~\cite{Berger:2001xd}, wide-angle Compton scattering (WACS)~\cite{Radyushkin:1998rt,Diehl:1998kh}, and double DVCS~\cite{Deja:2023ahc}.

Major experimental efforts have significantly advanced our knowledge of GPDs, with contributions from Hall A~\cite{JeffersonLabHallA:2006prd, JeffersonLabHallA:2007jdm} and CLAS~\cite{CLAS:2001wjj, CLAS:2006krx, CLAS:2007clm} at Jefferson Lab, ZEUS~\cite{ZEUS:1998xpo, ZEUS:2003pwh} and H1~\cite{H1:2001nez, H1:2005gdw} at HERA, HERMES~\cite{HERMES:2001bob, HERMES:2006pre, HERMES:2008abz} at DESY, and COMPASS~\cite{dHose:2004usi} at CERN, among others.

GPDs provide vital information about the internal structure of the proton, including the spatial distribution, spin contributions, and orbital angular momentum (OAM) of quarks and gluons. As a result, the precise determination of proton GPDs remains a central goal for future facilities such as the Electron-Ion Colliders (EICs)~\cite{Accardi:2012qut,AbdulKhalek:2021gbh,Anderle:2021wcy,AbdulKhalek:2022hcn,Abir:2023fpo,Amoroso:2022eow,Hentschinski:2022xnd}, the Large Hadron-Electron Collider (LHeC)~\cite{LHeCStudyGroup:2012zhm,LHeC:2020van}, and the 22 GeV upgrade at Jefferson Lab~\cite{Dudek:2012vr, Burkert:2018nvj,Accardi:2023chb}.

From the theoretical perspective, a range of nonperturbative methods has been developed to investigate quark GPDs, often within phenomenological frameworks~\cite{Pasquini:2005dk,Pasquini:2006dv,Meissner:2009ww,Boffi:2002yy,Scopetta:2003et,Choi:2001fc,Choi:2002ic,Mondal:2015uha,Chakrabarti:2015ama,Kaur:2023lun}. Among the most promising first-principles approaches is Euclidean lattice QCD, which has been extensively applied to extract information on GPDs~\cite{Ji:2013dva,Ji:2020ect,Lin:2021brq,Lin:2020rxa,Bhattacharya:2022aob, Alexandrou:2021bbo, Alexandrou:2022dtc, Guo:2022upw, Alexandrou:2020zbe, Gockeler:2005cj, QCDSF:2006tkx, Alexandrou:2019ali,Hannaford-Gunn:2024aix}. Another promising first principles approach is basis light-front quantization (BLFQ)~\cite{Vary:2009gt}, whose development and application to the proton is the focal point of this review.

While the study of gluon GPDs remains less developed compared to their quark counterparts, recent efforts have begun to address both chiral-even and chiral-odd gluon GPDs. These studies employ diverse approaches, including light-cone spectator models~\cite{Tan:2023kbl,Chakrabarti:2024hwx,Chakrabarti:2023djs}, BLFQ~\cite{Lin:2023ezw,Lin:2024ijo}, light-front holography (LFH)~\cite{Gurjar:2022jkx,deTeramond:2021lxc}, the double distribution representation~\cite{Goloskokov:2024egn}, and holographic string-based methods~\cite{Mamo:2024jwp,Mamo:2024vjh}, offering valuable insights into the leading-twist structure of gluons inside the proton.

Recently, numerous theoretical and experimental efforts~\cite{ZEUS:2003pwh,COMPASS:2008isr,HERMES:2009lmz,Bacchetta:2025ara,Bacchetta:2024yzl} have been devoted to understanding TMDs.
Unlike collinear PDFs, TMDs are sensitive to the intrinsic transverse motion of partons and their spin-dependent correlations, offering a more complete picture of hadron structure, particularly its transverse dynamics~\cite{Barone:2001sp,Collins:2011zzd,Rogers:2015sqa}. These distributions are essential for describing cross sections in high-energy scattering processes involving hadrons. According to QCD factorization theorems~\cite{Collins:1985ue,Collins:1987pm,Collins:1996fb,Sterman:1995fz,Collins:1998rz,Diehl:2003ny,Collins:2011zzd,Diehl:2011yj,Rogers:2015sqa}, TMDs particularly those of quarks can be extracted from SIDIS~\cite{Bacchetta2007,Ji2005} and Drell–Yan processes~\cite{Tangerman:1994eh,Collins:2002kn,Zhou:2009jm}.

A variety of QCD-inspired models have been employed to study nucleon TMDs, including the MIT bag model~\cite{Avakian2010}, covariant parton model~\cite{Efremov2009,Bastami2021a}, spectator model~\cite{Bacchetta2008,Bacchetta2020a,Bacchetta2021,Bacchetta2022}, light-front quark–diquark models motivated by soft-wall AdS/QCD~\cite{Maji2017a,Gurjar:2022rcl,Gurjar:2023uho,Gurjar:2024krn}, light-cone constituent models~\cite{Pasquini2008}, and the BLFQ framework~\cite{Hu:2022ctr,Zhu:2024awq,Yu:2024mxo}. In addition, first-principles approaches such as lattice QCD in discretized Euclidean space-time~\cite{Musch2011,Musch2012,Ji:2014hxa,Yoon2017,Constantinou:2020hdm} and the Dyson–Schwinger equations framework~\cite{Shi:2018zqd,Shi:2020pqe} have shown promise. However, Euclidean-space formulations face intrinsic challenges in determining TMDs directly.

In theoretical treatments of high-energy scattering, cross sections are organized as a power series in $1/Q$. The leading contribution comes from leading-twist (twist-2) distributions, which, within the parton model, admit a probabilistic interpretation as the density of partons inside a hadron~\cite{Jaffe:1983hp,Barone:2001sp,Collins:2011zzd,Aybat:2011zv}. Subleading contributions, suppressed by $1/Q$, involve twist-3 distributions. Unlike twist-2 distributions, which reflect single-parton probabilities, twist-3 TMDs encode multiparton correlations within the hadron~\cite{Jaffe:1991kp,Jaffe:1991ra,Efremov:2002qh,AbdulKhalek:2021gbh,Zhu:2024awq}.

A range of theoretical approaches have been developed to investigate the partonic structure of hadrons. Notably, lattice QCD calculations in Euclidean space-time~\cite{Hagler:2009ni,Joo:2019byq,MILC:2009mpl,BMW:2008jgk} and continuum methods based on DSEs~\cite{Maris:2003vk,Roberts:1994dr,Bashir:2012fs} provide first-principles insights into nonperturbative QCD. An alternative first principles framework is LF quantization of QCD~\cite{Brodsky:1997de}, which allows for a direct description of hadron structure in terms of partonic degrees of freedom. Within this approach, LFH offers a semiclassical perspective on confinement and dynamical symmetry breaking~\cite{Brodsky:2014yha}. A particularly successful Hamiltonian method is BLFQ, which enables nonperturbative and relativistic solutions of quantum field theoretical bound-state problems~\cite{Vary:2009gt,Zhao:2014xaa,Nair:2022evk,Wiecki:2014ola,Li:2015zda,Jia:2018ary,Lan:2019vui,Mondal:2019jdg,Xu:2021wwj,Lan:2021wok,Xu:2022yxb,Xu:2024sjt,Mondal:2025fdl}. 

In this review, we report recent progress in studying nucleon structure using the BLFQ framework.
Over the past several years, the BLFQ framework has undergone notable development in its application to nucleon structure, yielding increasingly sophisticated and realistic descriptions. Initial studies employed the leading Fock sector $|qqq\rangle$ within an effective LF Hamiltonian that incorporates a confining potential and one-gluon exchange, successfully reproducing key nucleon observables~\cite{Mondal:2019jdg,Xu:2021wwj}. Subsequent extensions included the next-to-leading Fock sector $|qqqg\rangle$, with the Hamiltonian refined to incorporate QCD-motivated interactions and three-dimensional confinement for the valence sector~\cite{Xu:2022yxb}. This advancement has enabled detailed studies of gluonic contributions to the nucleon's internal structure, including the gluon helicity and orbital angular momentum components relevant to proton spin decomposition, as well as its gravitational structure, mass decomposition, and three-dimensional imaging through GPDs and TMDs~\cite{Xu:2022yxb,Nair:2025sfr,Lin:2024ijo,Lin:2023ezw,Zhang:2025nll,Yu:2024mxo,Zhu:2024awq}. Most recently, the framework has achieved a milestone by solving for the nucleon  light-front wave functions (LFWFs) as eigenstates of the QCD Hamiltonian without introducing an explicit confining potential. These fully relativistic, nonperturbative calculations include the $|qqq\rangle$, $|qqqg\rangle$, and $|qqqq\bar{q}\rangle$ sectors, enabling predictions of quark and gluon matter densities, helicity and transversity distributions, and spin-related observables that show promising agreement with experimental and phenomenological results~\cite{Xu:2024sjt,Mondal:2025bjn}. These developments mark important steps toward a first-principles understanding of nucleon structure within the BLFQ approach.


\section{Light-Front Quantization}\label{LFQ} 
LF quantization provides a natural framework for describing the partonic structure of QCD at high energies. This formalism underpins many phenomenological studies of hard inclusive and exclusive processes. In this section, we discuss nonperturbative approaches to light-front Hamiltonian quantization, with a focus on the  QCD Hamiltonian formulated in the light-cone gauge (for comprehensive reviews, see Refs.~\cite{Brodsky:1997de,Hiller:2016itl}). We present the methods used to solve these Hamiltonians and highlight results for the nucleon. Particular emphasis is placed on Discretized Light-Cone Quantization (DLCQ) and BLFQ, as both frameworks enable the dynamical inclusion of gluons and sea quarks.
In these approaches,  symmetries and general properties of QCD are used to construct effective partonic amplitudes or distributions on the LF.

Before introducing specific methods for solving LF Hamiltonians, we briefly review the foundational concepts of the LF approach, which originate from Dirac’s formulation of Poincaré-invariant quantum dynamics~\cite{Dirac:1949cp}. The LF variables are defined in relation to equal-time coordinates as  
$P = (P^0 + P^3, P^0 - P^3, \mathbf{P}_\perp) = (P^+, P^-, \mathbf{P}_\perp)$,
where $P$ is the hadron's four-momentum.

For the hadron’s constituents—quarks, antiquarks, and gluons, collectively referred to as partons—we define $\bm{p}_{i\perp}$ as the transverse momentum of the $i$-th parton, $x_i = p_i^+/P^+$ as its longitudinal momentum fraction, and $\lambda_i$ as its LF  helicity~\cite{Soper:1972xc}.

The mass-squared eigenstates and their corresponding LFWFs are obtained by solving the LF Schrödinger equation. Setting $P^\perp = 0$ and identifying the LF Hamiltonian as $H_{\rm LF} = P^+P^- $, the eigenvalue equation reads:
\begin{equation}\label{eq:schrodinger}
    H_{\rm LF} |{P,\Lambda}\rangle = M^2 |{P,\Lambda}\rangle\,,
\end{equation}
where \( \Lambda \) denotes the hadron’s LF helicity and $M$ its mass. The Hamiltonian includes kinetic and interaction terms:
\begin{equation}\label{eq:schrodingerterms}
    H_{\rm LF} = \sum_i \frac{m_i^2 + \bm{p}_{i\perp}^2}{x_i} + H_{\rm int} \,.
\end{equation}
The sum runs over all partons, with $m_i $ representing the mass of the $i$-th parton. 

We note that the LF Hamiltonian eigenvalue problem applies to systems with arbitrary baryon number and is therefore applicable not only to individual hadrons but also to nuclei. A general eigenstate of the system can be expanded in a Fock-space basis as a sum over sectors with $N$ partons:
\begin{align}
        \ket{P,\Lambda}&=\sum_N \sum_{\lambda_1, . . .,\lambda_N} \int \frac{\prod_{i=1}^N  \mathrm{d} x_i \mathrm{d}^2 \bm{p}_{i\perp} }{\left[2(2\pi)^N\right]^2\sqrt{x_1 \dots x_N}} \delta\left(1-\sum_{i=1}^N x_i\right) \nonumber\\
        &\times\delta^2\left(\sum_{i=1}^N \bm{p}_{i\perp}\right) \psi^{\Lambda}_{\{\lambda_i\}_N}(\{p_i\}_N) \ket{\{\lambda_i,p_i\}_N} \, ,
\label{eq:Overall_LF_State_Vector}
\end{align}
where $\psi^{\Lambda}_{\{\lambda_i\}_N}(\{p_i\}_N)$ is the LF helicity amplitude, and $\ket{\{\lambda_i,p_i\}_N}$ denotes a properly normalized $N$-parton state consisting of quark, antiquark, and gluon creation operators acting on the vacuum. This expression is schematic: for each $N$, multiple configurations may contribute, depending on the specific partonic content, even with the same net fermion number. 
In the following sections, we introduce two related approaches: DLCQ and BLFQ that provide practical realizations of Eq.~\eqref{eq:Overall_LF_State_Vector} for numerical computation.

For gauge theories, it is standard to adopt the LF gauge, $A^+ = 0$, which allows for the elimination of non-dynamical degrees of freedom via constraint equations. In both QED and QCD, this gauge fixing yields interaction Hamiltonians $H_{\rm int}$ composed of Pauli spinors and a minimal set of boson–fermion (for QED and QCD) and boson–boson (for QCD) vertices. Additionally, instantaneous interactions arise due to the elimination of constrained fields, leading to higher-order terms that exhibit divergences. 


As in the Lagrangian formalism, Hamiltonian field theory requires both regularization and renormalization. While dimensional regularization is well-suited for perturbative analyses, nonperturbative formulations typically rely on other methods such as invariant mass cutoffs or Pauli–Villars regularization. Additionally, numerical solutions to nonperturbative eigenvalue problems necessitate finite discretization. In some approaches, discretization itself serves as a form of regularization, as seen in DLCQ and BLFQ. Alternatively, discretization may be used strictly as a numerical tool, in which case the continuum limit must be carefully taken. Owing to the kinematical nature of LF boosts, cluster decomposition remains valid even in the continuum, allowing renormalization techniques akin to those used in perturbation theory to be extended. A notable realization of this idea is the Fock Sector Dependent Renormalization (FSDR) method~\cite{Karmanov:2008br}.

Another important nonperturbative framework is the Similarity Renormalization Group (SRG), which is grounded in Wilsonian renormalization group evolution~\cite{Glazek:1993rc, Wilson:1994fk}. Benefiting from QCD’s property of asymptotic freedom, the SRG transformation can be computed perturbatively up to a few GeV. Various formulations of SRG have been developed, including the Bloch–Wilson approach~\cite{Perry:1993mn} and the Renormalization Group Procedure for Effective Particles (RGPEP)~\cite{Glazek:2012qj}. Within RGPEP, effective Hamiltonians incorporating a gluon mass ansatz have been constructed for heavy-flavor systems~\cite{Glazek:2017rwe}, and the formalism has been shown to reproduce asymptotic freedom in the three-gluon vertex~\cite{Gomez-Rocha:2015esa}.

The Fock space expansion of the eigenstate, given in Eq.~\eqref{eq:Overall_LF_State_Vector}, provides a natural basis for solving the LF Hamiltonian eigenvalue equation, Eq.~\eqref{eq:schrodinger}. This expansion leads to an infinite hierarchy of coupled integral equations, which must be truncated for practical computation. This structure is reminiscent of the Dyson–Schwinger and Bethe–Salpeter equations in covariant formulations. The Light-Front Tamm–Dancoff Approximation (LFTDA) achieves this truncation by limiting the Fock space to a finite number of partons~\cite{Perry:1990mz}. When combined with FSDR~\cite{Karmanov:2008br}, LFTDA has been successfully applied to a variety of field theories, as reviewed in Ref.~\cite{Hiller:2016itl}. While convergence can typically be tested numerically~\cite{Li:2015iaw}, the computational complexity grows rapidly with the inclusion of higher Fock sectors. To mitigate issues arising from truncation, the Light-Front Coupled Cluster (LFCC) method has been proposed, offering an alternative based on a coherent-state expansion~\cite{Chabysheva:2011ed}.

A further major advancement in LF quantization is the development of  LFH, which establishes a correspondence between LFQCD, conformal quantum mechanics, and gravity in a higher-dimensional anti-de Sitter (AdS) space. This approach leads to a remarkably simple and universal confining potential and provides valuable insights into hadronic structure~\cite{Brodsky:2014yha}.
\subsection{Discretized Light-Cone Quantization}
While lattice QCD calculations solve QCD in Euclidean spacetime, DLCQ addresses the problem directly in Minkowski spacetime by discretizing the momentum space basis (see Ref.~\cite{Brodsky:1997de} and references therein).

In DLCQ, the momentum space is discretized by imposing boundary conditions, either periodic or anti-periodic, on standing waves within a finite box of length $L$ in each transverse and longitudinal direction. Early applications of DLCQ to gauge theories demonstrated its potential, including the study of positronium in strongly coupled QED~\cite{Krautgartner:1991xz} and the successful solution of QCD in $(1+1)$-dimensions~\cite{Hornbostel:1988fb}. Extensions to 3+1 dimensional QCD using DLCQ revealed both formal and computational challenges, but nonetheless yielded valuable insights, as reviewed in Ref.~\cite{Hiller:2016itl}.

A hybrid formulation combining DLCQ with lattice methods has also been explored to compute PDFs for selected meson states across a range of couplings~\cite{Burkardt:2001jg, Chakrabarti:2003wi}. These developments motivated the search for approaches that preserve LF kinematic symmetries while enhancing computational efficiency, paving the way for further advances in LF Hamiltonian methods.

\subsection{Basis Light-Front Quantization}\label{sec:BLFQ}
The development of LF Hamiltonian approaches in Minkowski space, with full preservation of kinematic symmetries, began with the use of basis function methods to solve LF wave equations~\cite{Harindranath:1992pk}. This effort culminated in the introduction of the BLFQ framework~\cite{Vary:2009gt}, designed to address gauge theory Hamiltonians using basis functions that satisfy general mathematical criteria and respect LF kinematic symmetries. Moreover, BLFQ is particularly well-suited for the long-term objective of constructing basis functions that incorporate key dynamical features of QCD, such as confinement and chiral symmetry breaking, thereby enhancing convergence in nonperturbative LF QCD calculations.  A brief review of BLFQ and its applications to mesons and baryons prior to 2 years ago can be found in Ref.~\cite{Gross:2022hyw} 

In BLFQ, the representation of LF eigenstates differs from the momentum-space formulation given in Eq.~\eqref{eq:Overall_LF_State_Vector}. Instead of relying on LF plane waves, the approach constructs eigenstates as superpositions of orthonormal $N$-parton Fock-space states, where each parton occupies a mode from a chosen single-particle basis. This replaces the conventional LF plane-wave quantization with quantization in terms of modes that solve a single-parton LF Schrödinger equation akin to Eq.~\eqref{eq:schrodinger}. Consequently, the many-parton basis states are built from fermion, antifermion, and boson creation operators acting on the vacuum, with each operator populating an independent mode of the single-parton basis.

For practical applications, the transverse modes are typically taken from the two-dimensional harmonic oscillator (2D-HO) basis, which preserves transverse boost invariance and aligns naturally with insights from holographic LFQCD~\cite{Brodsky:2014yha}. For the longitudinal direction, various choices have been employed such as DLCQ and a basis related to Jacobi polynomials~\cite{Li:2015zda,Li:2017mlw}. In principle, the basis selection is flexible, subject to general mathematical constraints, with computational convenience and numerical efficiency guiding specific implementations.

Each single-parton mode is denoted by a lowercase Greek letter, which encapsulates the complete set of quantum numbers such as spatial, spin, color, and flavor degrees of freedom associated with that mode. The single-parton states for fermions and bosons form an orthonormal and complete basis, with their respective creation operators satisfying standard anticommutation or commutation relations.

Within the BLFQ framework, a physical eigenstate is expressed as a Fock-space expansion over sectors with $N$ partons:
\begin{gather}
    \scalebox{1.0}{$\begin{aligned}
        \ket{P,\Lambda}&=\sum_N \sum_{\{\alpha_i\}_N}  \psi^{\Lambda}_{\{\alpha_i\}_N} 
        \ket{\{\alpha_i\}_N} \, .
    \end{aligned}$}\label{eq:Overall_BLFQ_State_Vector}
\end{gather}
where the inner sum runs over all valid $N$-parton configurations that satisfy global symmetries such as baryon number, electric charge, total LF momentum, helicity projection along the $x^-$ direction, and flavor conservation. For configurations involving two or more identical bosons in the same mode, an additional symmetry factor ensures proper normalization.

At this stage, the Hamiltonian eigenvalue problem of Eq.~\eqref{eq:schrodinger} remains infinite-dimensional due to the unbounded number of Fock sectors and single-parton modes. However, with  appropriately chosen QCD interaction vertices, one can obtain meaningful predictions for low-resolution observables including hadron spectroscopy, electroweak transitions, form factors at low $Q^2$, and parton distributions using practical truncations in both mode space and Fock space. 

The flexibility and robustness of BLFQ have been demonstrated through its successful applications in both QED and QCD. In QED, the framework has been used to compute the electron’s anomalous magnetic moment with high precision~\cite{Zhao:2014xaa,Honkanen:2010rc}, as well as GPDs~\cite{Chakrabarti:2014cwa}, TMDs~\cite{Hu:2020arv}, the structure of real and virtual photons~\cite{Nair:2022evk,Nair:2023lir}, and the strongly bound positronium system~\cite{Wiecki:2014ola}. In QCD, BLFQ has been applied to a wide range of hadronic systems, including light mesons~\cite{Jia:2018ary,Lan:2019vui,Lan:2019rba,Lan:2021wok,Adhikari:2021jrh,Mondal:2021czk,Wu:2024hre,Zhu:2023lst,Kaur:2024iwn,Lan:2025fia,Lan:2024ais}, heavy quarkonia~\cite{Li:2015zda,Li:2017mlw,Li:2018uif,Lan:2019img}, and heavy-light mesons~\cite{Tang:2018myz,Tang:2019gvn}. More recently, the approach has been extended to baryonic systems such as the proton~\cite{Mondal:2019jdg,Xu:2021wwj,Du:2019ips,Kaur:2023lun,Zhang:2023xfe,Peng:2024qpw,Hu:2022ctr,Liu:2022fvl,Nair:2024fit,Xu:2022yxb,Zhu:2024awq,Yu:2024mxo,Lin:2023ezw,Lin:2024ijo,Zhang:2025nll,Nair:2025sfr,Xu:2024sjt}, the $\Lambda$ and $\Lambda_c$ baryons, and their isospin triplet partners: $\Sigma^0$, $\Sigma^+$, $\Sigma^-$, and $\Sigma_c^0$, $\Sigma_c^+$, $\Sigma_c^{++}$~\cite{Peng:2022lte,Zhu:2023nhl}, as well as the $\Lambda_b$ and its isospin triplet partners: $\Sigma_b^0$, $\Sigma_b^+$, and $\Sigma_b^{++}$~\cite{Meng:2025sys}. These studies provide insights into parton distribution functions, form factors, spin and mass decompositions, and three-dimensional imaging through GPDs and TMDs. Together, these applications underscore the growing capability of BLFQ as a first-principles tool for understanding the nonperturbative structure of matter in quantum field theory.

\section{BLFQ Approach to the Nucleon}\label{sec:BLFQ_application}
The internal structure of the nucleon is encapsulated in its LFWFs, which are obtained by solving the LF Hamiltonian eigenvalue equation, Eq.~\eqref{eq:schrodinger}. At fixed LF time ($x^+=t+z=0$), the nucleon state can be expressed through a Fock-space expansion of the schematic form:
\begin{align}\label{eq:Fock_space}
|\Psi\rangle =\,& \psi^{(3q)}|qqq\rangle + \psi^{(3q+g)}|qqqg\rangle + \psi^{(3q+u\bar{u})}|qqqu\bar{u}\rangle \nonumber\\
& + \psi^{(3q+d\bar{d})}|qqqd\bar{d}\rangle + \psi^{(3q+s\bar{s})}|qqqs\bar{s}\rangle + \dots\,,
\end{align}
where each $\psi^{(\dots)}$ denotes the probability amplitude for a specific partonic configuration, defining the LFWFs, which can be formulated in either coordinate or momentum space. These LFWFs serve as the foundation for calculating a wide range of observables related to the nucleon's internal structure. Within the BLFQ framework, each Fock sector comprises an infinite set of basis states. For practical numerical computations, applications to date implement both a truncation of the Fock space and a cutoff on the basis states within each Fock sector.

The basis states for each Fock particle are characterized by their longitudinal and transverse coordinates, as well as their helicity quantum numbers~\cite{Zhao:2013cma}. In the longitudinal direction, we follow DLCQ and a plane-wave basis confined within a one-dimensional box of length  $2L$, imposing antiperiodic boundary conditions for quarks and periodic boundary conditions for gluons. Consequently, the amplitude in longitudinal coordinate space takes the form:
\begin{eqnarray}
\Psi_k(x^-) = \frac{1}{2L} e^{i\frac{\pi}{L}kx^-}.
\end{eqnarray} 
The longitudinal momentum is discretized as 
$p^+ = \frac{2\pi}{L}k$,
where the dimensionless momentum index \( k \) takes half-integer values 
\( \frac{1}{2}, \frac{3}{2}, \frac{5}{2}, \ldots \) for fermions, and integer values 
\( 1, 2, 3, \ldots \) for bosons. The bosonic zero mode is omitted. 
All many-body basis states are constructed to have the same total longitudinal momentum,
$P^+ = \sum_i p_i^+$,
where the sum runs over all particles in a given basis state. 
We introduce a dimensionless variable \( K = \sum_i k_i \) to parameterize the total longitudinal momentum as 
$P^+ = \frac{2\pi}{L}K$.
For an individual particle \( i \), the longitudinal momentum fraction is then given by  
\begin{align}
x_i = \frac{p_i^+}{P^+} = \frac{k_i}{K}.
\end{align}

In the transverse direction, we employ the two-dimensional harmonic oscillator (2D-HO) basis \(\phi_{n m}(\bm{p}_{\perp}; b)\), which is characterized by the quantum numbers \(n\) and \(m\), corresponding to the radial excitation and the angular momentum projection, respectively. In momentum space, the orthonormalized 2D-HO wave functions are given by~\cite{Vary:2009gt,Zhao:2013cma}
\begin{align}
\phi_{n,m}(\bm{p}_{\perp};b)
 =&\frac{\sqrt{2}}{b(2\pi)^{\frac{3}{2}}}\sqrt{\frac{n!}{(n+|m|)!}}e^{-\bm{p}_{\perp}^2/(2b^2)}\left(\frac{|\bm{p}_{\perp}|}{b}\right)^{|m|}L^{|m|}_{n}\left(\frac{\bm{p}_{\perp}^2}{b^2}\right)e^{im\theta},\label{ho}
\end{align}
where \(b\) is the HO basis scale parameter with dimensions of mass, and \(\bm{p}_{\perp}\) denotes the transverse momentum of the particle. The functions \(L^{\alpha}_{n}(x)\) are the generalized Laguerre polynomials, and \(\theta = \arg(\bm{p}_{\perp}/b)\) is the azimuthal angle. 

For spin degrees of freedom, the quantum number \(\lambda\) labels the helicity of the particle. Each single-parton basis state is therefore labeled by a set of four quantum numbers: \(\bar{\alpha} = \{k, n, m, \lambda\}\). Furthermore, we require that the many-body basis states possess a well-defined total angular momentum projection:
\begin{align}
M_J = \sum_i \left(m_i + \lambda_i\right).
\end{align}

In Fock sectors that admit multiple color-singlet configurations, additional quantum labels are introduced to distinguish among them. For example, the 
$|qqqg\rangle$ sector contains two independent color-singlet combinations, while the 
$|qqqq\bar{q}\rangle$ sector contains three. The many-parton basis states in each Fock sector are constructed as direct products of ordered configurations in flavor, spin, and spatial degrees of freedom, combined with global color-singlet states. This is adequate for full antisymmetrization of the valence Fock sector. Future developments aim to refine this construction by enforcing total antisymmetry more precisely in the Fock sectors beyond the valence sector. With this basis, the LF Hamiltonian eigenvalue problem is transformed into a finite-dimensional matrix eigenvalue problem, suitable for numerical diagonalization.

To make the Hamiltonian matrix finite and numerically tractable, we introduce two truncation parameters: \(N_{\rm max}\) and \(K\)~\cite{Zhao:2014xaa}. The transverse basis is truncated using \(N_{\rm max}\), which limits the total 2D-HO excitation energy of all partons according to
\begin{align}
\sum_i \left( 2 n_i + |m_i| + 1 \right) \leq N_{\rm max}.
\end{align}
This cutoff acts as an effective regulator of the transverse structure of the LFWFs and ensures the factorization of the center-of-mass motion within the overcomplete single-parton basis space. It introduces both infrared (IR) and ultraviolet (UV) cutoffs in the transverse direction, approximately given by
\begin{align}
\Lambda_{\rm IR} \sim \frac{b}{\sqrt{N_{\rm max}}}
\quad \text{and} \quad
\Lambda_{\rm UV} \sim b \sqrt{N_{\rm max}}.
\end{align}

The longitudinal basis is truncated using the parameter \(K\), which restricts
the longitudinal resolution of the basis and directly determines the resolution of the PDFs.

The LFWFs of the nucleon with helicity $\Lambda$ in momentum space are written as a sum over contributions from individual Fock sectors. Each wave function is expressed as:
\begin{equation}
\Psi^{{N},\,\Lambda}_{\{x_i,\bm{p}_{i\perp},\lambda_i\}} = \sum_{\{n_i m_i\}} \psi^{{N}}(\{\overline{\alpha}_i\}) \prod_{i=1}^{{N}} \phi_{n_i m_i}(\bm{p}_{i\perp}, b)\,,
\label{eqn:wf}
\end{equation}
where \(\psi^{{N}}(\{\overline{\alpha}_i\})\) denotes the LF amplitude for a Fock sector with \({N}\) partons, obtained by diagonalizing the full Hamiltonian matrix in the BLFQ basis. 

\begin{figure}
\begin{center}
\includegraphics[width=1.0\linewidth]{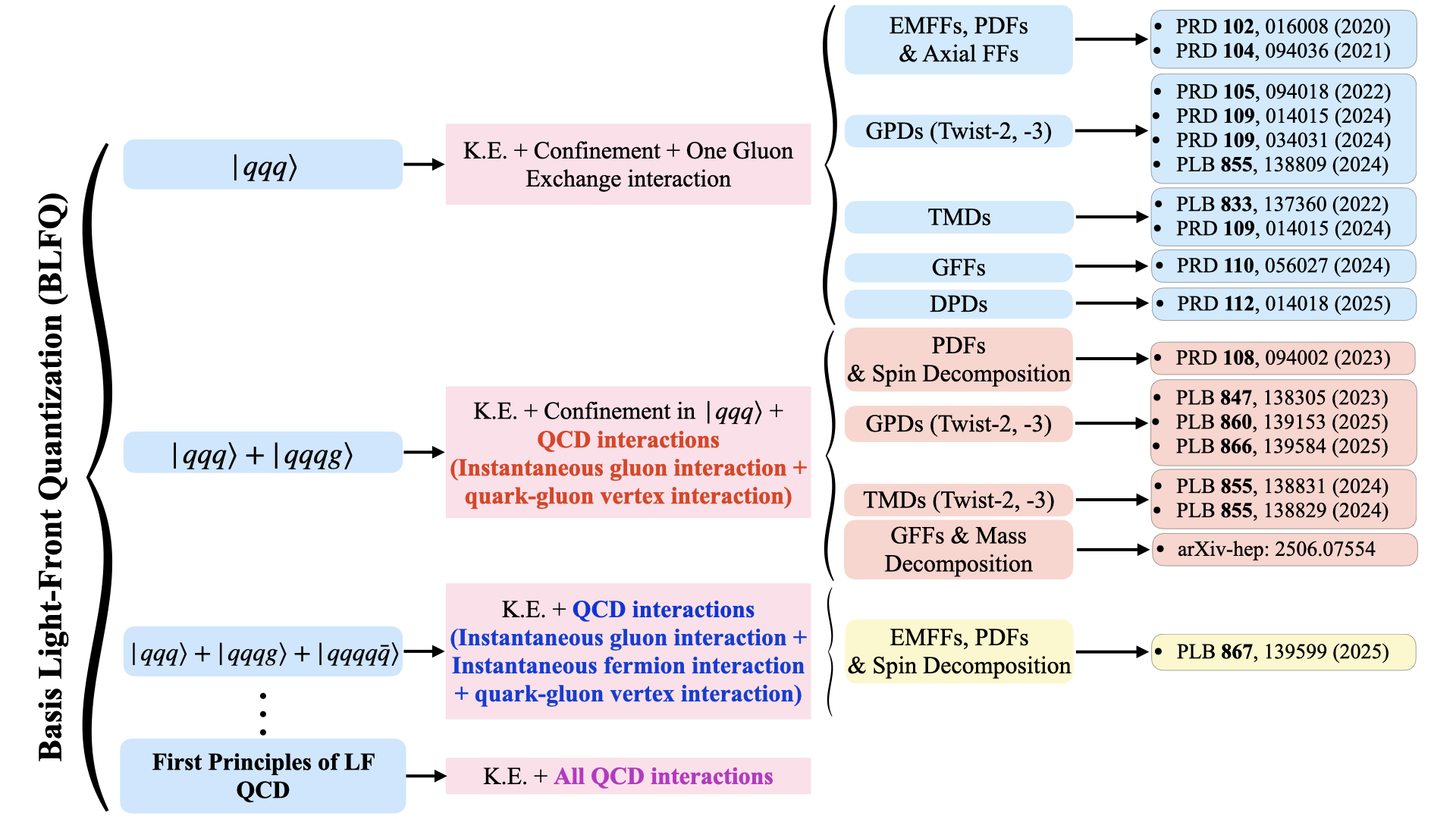}
\caption{Overview of recent advances in applying BLFQ to nucleon structure studies with different Fock-sector components, including calculations of FFs, PDFs, GPDs, TMDs, and other key observables.}
\label{publications}
\end{center}
\end{figure}

Although LFWFs are expected to respect parity symmetry ($\mathcal{P}$), this symmetry is broken by Fock-space truncation. However, one can instead employ mirror parity \(\hat{P}_x = \hat{R}_x(\pi) \mathcal{P}\)~\cite{Brodsky:2006ez} as an effective replacement. Under mirror parity, the wave functions satisfy the following relation:
\begin{align}
\Psi^{\downarrow}_{\{x_i,n_i,m_i,\lambda_i\}} = (-1)^{\sum_i m_i + 1} \Psi^{\uparrow}_{\{x_i,n_i,-m_i,-\lambda_i\}}\,,
\end{align}
where the arrows indicate the helicity of the nucleon. We summarize recent advances in applying BLFQ to nucleon structure studies in Fig.~\ref{publications}.

\subsection{Nucleon Structure with Valence Fock Component}
Initial applications of BLFQ to the nucleon focused on the leading Fock sector $|qqq\rangle$ using an effective LF  Hamiltonian that combined a confining potential with one-gluon exchange (OGE). This approach provided initial successes for key nucleon observables. The high-precision results from BLFQ treatments of QED problems~\cite{Wiecki:2014ola,Zhao:2014xaa} offer a pathway to model the OGE interaction ($H_{\text{OGE}}$) between fermions in QCD, which governs the dominant short-distance physics in hadrons. For long-distance dynamics, the confining interaction is derived from LFH, supplemented by a suitable longitudinal confinement term, collectively forming $H_{\text{conf}}$. The total effective LF interaction is then given by $H_{\text{int}} = H_{\text{conf}} + H_{\text{OGE}}$.

Analogous to the nuclear shell model, the solvable part of the Hamiltonian can be defined as the sum of the kinetic energy and the confining interaction, $H_0 = H_{\text{kin}} + H_{\text{conf}}$. This choice establishes the zeroth-order Hamiltonian for implementing LFH, augmented with longitudinal confinement.

For the valence Fock sector of the nucleon, we adopt a light-front effective Hamiltonian $H_{\text{eff}} (= P^- P^+)$ given by
\begin{align}\label{eq:hamiltonian}
H_{\text{eff}} = \sum_{i} \frac{m_i^2 + \bm{p}_{i\perp}^2}{x_i} 
+ \frac{1}{2}\sum_{i\ne j} V_{i,j}^{\text{conf}} 
+ \frac{1}{2}\sum_{i\ne j} V_{i,j}^{\text{OGE}},
\end{align}
where $\sum_i x_i = 1$, $m_i$ is the constituent quark mass, and $i,j$ label particles within a Fock sector. 

The confining potential $V^{\text{conf}}_{i,j}$ incorporates both transverse and longitudinal confinement. We generalize the soft-wall holographic confinement~\cite{Brodsky:2014yha} in the transverse direction and include a longitudinal confining potential that reproduces three-dimensional confinement in the nonrelativistic limit~\cite{Li:2015zda}. For a many-body system, the complete confining potential takes the form
\begin{align}\label{conf}
V^{\text{conf}}_{i,j} = \kappa_T^4 \bm{r}_{ij\perp}^2 
+ \frac{\kappa_L^4}{(m_i + m_j)^2} \partial_{x_i}(x_i x_j \partial_{x_j}),
\end{align}
where $\bm{r}_{ij\perp} = \sqrt{x_i x_j}(\bm{r}_{i\perp} - \bm{r}_{j\perp})$ is the transverse relative coordinate, $\kappa_T$ ($\kappa_L$) denotes the transverse (longitudinal) confinement strength, and $\partial_x \equiv (\partial/\partial x)_{\bm{r}_{ij\perp}}$.

The OGE interaction in Eq.~\eqref{eq:hamiltonian} is given by
\begin{align}
V^{\text{OGE}}_{i,j} = \frac{4\pi C_F \alpha_s}{Q^2_{ij}} 
\bar{u}_{s'_i}(p'_i)\gamma^\mu u_{s_i}(p_i)
\bar{u}_{s'_j}(p'_j)\gamma_\mu u_{s_j}(p_j),
\end{align}
with fixed coupling constant $\alpha_s$. Here, $Q^2_{ij}$ represents the average squared four-momentum transferred by the gluon:
\begin{align}
Q^2_{ij} = \frac{1}{2} \Bigg[ &\left(\frac{\bm{p}_{i\perp}^2 + m_i^2}{x_i} - \frac{\bm{p}_{i\perp}^{\prime 2} + m_i^2}{x^{\prime}_i} - \frac{(\bm{p}_{i\perp}^2 - \bm{p}_{i\perp}^{\prime 2}) + \mu_g^2}{x_i - x^{\prime}_i}\right)\nonumber\\
&  - (i \rightarrow j) \Bigg],
\end{align}
where $\mu_g$ is the gluon mass. The color factor $C_F = -2/3$ indicates an attractive potential. The spinors $u_{s_i}(p_i)$ satisfy the free Dirac equation, with $s_i$ denoting the spin state and $\bm{p}_{i\perp}$ the transverse momentum of quark $i$. 

The OGE interaction naturally generates dynamical spin structure in the LFWFs, which is essential for computing spin-dependent observables.

In our approach, we construct the basis using single-particle coordinates. This choice offers two key advantages: (i) it treats all particles in the Fock space equivalently, and (ii) it facilitates handling symmetry operations for identical particles~\cite{Zhao:2014xaa}. However, the effective Hamiltonian $H_{\text{eff}}$ involves both the transverse center-of-mass (CM) motion and the intrinsic dynamics. 

To address this issue, we introduce a constraint term:
\begin{align}\label{eq:Hprime}
    H^{\prime} = \lambda_{L} \left( H_{\text{CM}} - 2b^2 I \right),
\end{align}
where $I$ is the identity operator and $\lambda_L$ serves as a Lagrange multiplier. The CM Hamiltonian is given by~\cite{Wiecki:2014ola}:
\begin{align}
    H_{\text{CM}} = \left(\sum_i \bm{p}_{i\perp}\right)^2 + b^4\left(\sum_i x_i \bm{r}_{i\perp}\right)^2.
\end{align}
By choosing $\lambda_L$ sufficiently large and positive, we elevate the excited CM states to higher energies while preserving the low-lying physical states. The complete effective Hamiltonian for diagonalization becomes:
\begin{align}
  H_{\text{eff}}^{\prime} = H_{\text{eff}} - \left(\sum_i \bm{p}_{i\perp}\right)^2 + \lambda_L \left(H_{\text{CM}} - 2b^2I\right).
\end{align}

Diagonalizing $H_{\text{eff}}^{\prime}$ within the BLFQ framework yields both the mass spectrum (eigenvalues) and the LFWFs encoded in the eigenvectors. The ground state is identified as the nucleon state, with its valence LFWF in momentum space expressed in our symmetry-preserving orthonormal basis as given in Eq.~\eqref{eqn:wf}.

The parameters of this effective Hamiltonian have been calibrated to reproduce both the nucleon mass and flavor electromagnetic form factors~\cite{Mondal:2019jdg,Xu:2021wwj}. 
The resulting LFWFs have proven remarkably successful in predicting diverse nucleon properties, including electromagnetic and axial form factors, charge and mass radii, PDFs, GPDs, TMDs, angular momentum distributions, gravitational form factors, and double parton distributions~\cite{Mondal:2019jdg,Xu:2021wwj,Liu:2022fvl,Hu:2022ctr,Nair:2024fit,Peng:2024qpw} (see Fig.~\ref{publications}).

\subsection{Nucleon Structure with One Dynamical Gluon}
BLFQ has successfully extended beyond the leading Fock component in QCD to solve for unflavored light mesons and nucleons with one dynamical gluon~\cite{Lan:2021wok,Xu:2022yxb}. At the initial scale, where baryons are described by both $|qqq\rangle$ and $|qqqg\rangle$ components, we consider the LF Hamiltonian:
\begin{equation}
P^- = P^-_{\text{QCD}} + P^-_C.
\end{equation}
Here $P^-_{\text{QCD}}$ represents the QCD Hamiltonian incorporating interactions relevant to these two leading Fock components, while $P^-_C$ models the confining interaction. In LF gauge with one dynamical gluon~\cite{Brodsky:1997de}:

\begin{align}
P_{\text{QCD}}^- = & \int d^2 x^{\perp}dx^- \Bigg\{\frac{1}{2}\bar{\psi}\gamma^+\frac{m_{0}^2+(i\partial^\perp)^2}{i\partial^+}\psi \nonumber\\
& + \frac{1}{2}A_a^i\left[m_g^2+(i\partial^\perp)^2\right] A^i_a \nonumber\\
& + g_s\bar{\psi}\gamma_{\mu}T^aA_a^{\mu}\psi \nonumber\\
& + \frac{1}{2}g_s^2\bar{\psi}\gamma^+T^a\psi\frac{1}{(i\partial^+)^2}\bar{\psi}\gamma^+T^a\psi\Bigg\},\label{eqn:PQCD}
\end{align}
where $\psi$ and $A^\mu$ denote the quark and gluon fields respectively, $T^a$ are the $SU(3)$ color generators, and $\gamma^\mu$ are Dirac matrices. The first two terms give the quark and gluon kinetic energies with bare masses $m_0$ and $m_g$ (where we introduce an effective gluon mass to fit nucleon form factors). The remaining terms represent the QCD vertex and instantaneous interactions with coupling $g_s$.

Following FSDR procedures developed for positronium~\cite{Zhao:2014hpa,Zhao:2020kuf} and mesons~\cite{Lan:2021wok}, we introduce a quark mass counterterm ($\delta m$) such that $m_0 = m_q + \delta m$, with $m_q$ being the physical quark mass. For the vertex interaction, we allow an independent quark mass $m_f$ following Ref.~\cite{Glazek:1992aq}, while neglecting identical quark antisymmetrization in this initial work.

Confinement in the leading Fock sector is implemented as~\cite{Li:2015zda}:
\begin{equation}
P_{\text{C}}^-P^+ = \frac{\kappa^4}{2}\sum_{i\ne j} \left\{\bm{r}_{ij\perp}^2 - \frac{\partial_{x_i}(x_i x_j\partial_{x_j})}{(m_i+m_j)^2}\right\}, \label{eqn:PC}
\end{equation}
where $\bm{r}_{ij\perp}=\sqrt{x_i x_j}(\bm{r}_{i\perp}-\bm{r}_{j\perp})$ is the holographic relative coordinate~\cite{Brodsky:2014yha}, $\kappa$ sets the confinement scale, and $\partial_x \equiv (\partial/\partial x)_{\bm{r}_{ij\perp}}$. Note that the confinement potential in Eq.~\eqref{conf} reduces to the form given in Eq.~\eqref{eqn:PC} when the transverse and longitudinal confinement strengths are set equal ($\kappa_T = \kappa_L = \kappa$).   We omit explicit $|qqqg\rangle$ sector confinement, relying on transverse basis limitations and the gluon mass to approximate confinement effects at this development stage, similar to functional approaches where gluons acquire effective masses~\cite{Cornwall:1981zr,Alkofer:2000wg,Deur:2016tte}.

Diagonalization yields mass spectra $M^2$ and momentum-space LFWFs (Eq.~\eqref{eqn:wf}), with $\psi^{{N}=3}(\{\overline{\alpha}_i\})$ and $\psi^{{N}=4}(\{\overline{\alpha}_i\})$ describing the $\ket{uud}$ and $\ket{uudg}$ sectors respectively. 

For nucleon calculations with one dynamical gluon, we use $N_{\text{max}}=9$ and $K=16.5$. Hamiltonian parameters are tuned to reproduce the proton mass and flavor form factors~\cite{Xu:2022yxb}, yielding $44\%$ probability for $|qqq\rangle$ and $56\%$ for $|qqqg\rangle$ at the model scale. These LFWFs successfully predict diverse proton observables including electromagnetic form factors, radii, parton distributions (PDFs, GPDs, TMDs), and spin/orbital angular momentum properties~\cite{Xu:2022yxb,Yu:2024mxo,Zhu:2024awq} (see Fig.~\ref{publications}).

\subsection{Nucleon Structure with Dynamical Gluons and Sea Quarks}
\label{subsec:dynamical}

Recent developments in BLFQ have successfully incorporated both dynamical gluon and sea quark degrees of freedom in nucleon structure calculations~\cite{Xu:2024sjt,Mondal:2025bjn} (see Fig.~\ref{publications}). The LF QCD Hamiltonian in the $A^+ = 0$ gauge, formulated for the three-quark ($|qqq\rangle$), three-quark-gluon ($|qqqg\rangle$), and three-quark-quark-antiquark ($|qqqq\bar{q}\rangle$) Fock sectors (see Eq.~\eqref{eq:Fock_space}), takes the form~\cite{Brodsky:1997de}:
\begin{align}\label{eqn:PQCD-sea}
P^-_{\text{QCD}} = & \int \mathrm{d}^2x^\perp \mathrm{d}x^- \Bigg\{ \frac{1}{2}\bar{\psi}\gamma^+\frac{(m_q+\delta m_q)^2+(i\partial^\perp)^2}{i\partial^+}\psi \nonumber \\
& + \frac{1}{2}A_a^\mu[\delta m_{g}^2+(i\partial^\perp)^2]A_\mu^a \nonumber \\
& + g_s\bar{\psi}\gamma^\mu T^aA^a_\mu\psi \nonumber \\
& + \frac{g_s^2}{2}\bar{\psi}\gamma^+T^a\psi\frac{1}{(i\partial^+)^2}\bar{\psi}\gamma^+T^a\psi \nonumber \\
& + \frac{g_s^2}{2}\bar{\psi}\gamma^\mu T^a A_\mu^a\frac{\gamma^+}{i\partial^+}\gamma^\nu T^b A_\nu^b\psi \Bigg\}.
\end{align}
The Hamiltonian contains several key components: the first two terms represent the kinetic energy of quarks (with mass $m_q$) and gluons (with zero bare mass), while the remaining terms correspond to vertex and instantaneous interactions governed by the strong coupling constant $g_s$. Notably, the final term represents an instantaneous fermion interaction that appears only in the three-Fock-component formulation, distinguishing it from the two-component case in Eq.~\eqref{eqn:PQCD}. To handle divergences and sector-dependent effects, we implement a FSDR scheme originally developed for positronium systems~\cite{Zhao:2014hpa,Zhao:2020kuf} and later extended to hadronic systems~\cite{Lan:2021wok,Xu:2022yxb}. This approach introduces quark and gluon mass counterterms ($\delta m_q$ and $\delta m_g$ respectively) and employs a separate quark mass parameter $m_f$ to model nonperturbative dynamics in vertex interactions~\cite{Burkardt:1998dd,Glazek:1992aq}.

This work represents significant progress beyond previous studies~\cite{Lan:2021wok,Xu:2022yxb} by eliminating phenomenological confinement potentials and using the physical gluon mass, $m_g=0$, while explicitly including sea quark contributions. The nucleon LFWFs in momentum space combine contributions from multiple Fock sectors, expressed as $\Psi = \sum_{{N}}\psi^{{N}}(\{\overline{\alpha}_i\})$, where $\psi^{{N}}$ represents the LF amplitude for sectors with ${N}$ partons, including valence ($|qqq\rangle$), dynamical gluon ($|qqqg\rangle$), and dynamical sea quark ($|qqqq\bar{q}\rangle$) components.

Numerical calculations employ basis truncation parameters $N_{\text{max}} = 7$ and $K = 16.5$, with a harmonic oscillator scale $b = 0.6$ GeV and an ultraviolet cutoff $b_{\text{inst}} = 2.80 \pm 0.15$ GeV for the instantaneous interaction. The model parameters $\{m_u, m_d, m_f, g_s\} = \{1.0, 0.85, 5.45 \pm 0.4, 2.90 \pm 0.1\}$ (in GeV units except for the dimensionless $g_s$) are determined by fitting the proton mass and electromagnetic properties~\cite{Xu:2024sjt,Mondal:2025bjn}. The large constituent quark masses in the $|qqq\rangle$ and $|qqqg\rangle$ sectors effectively mimic confinement effects, while a mass splitting of 0.15 GeV between up and down quarks accounts for isospin breaking. In the $|qqqq\bar{q}\rangle$ sector, we use current quark masses ($m_u = 2.2$ MeV, $m_d = 4.7$ MeV, $m_s = 94$ MeV) since higher Fock sectors are not included in the current model.

\begin{figure}
\begin{center}
\includegraphics[width=0.5\linewidth]{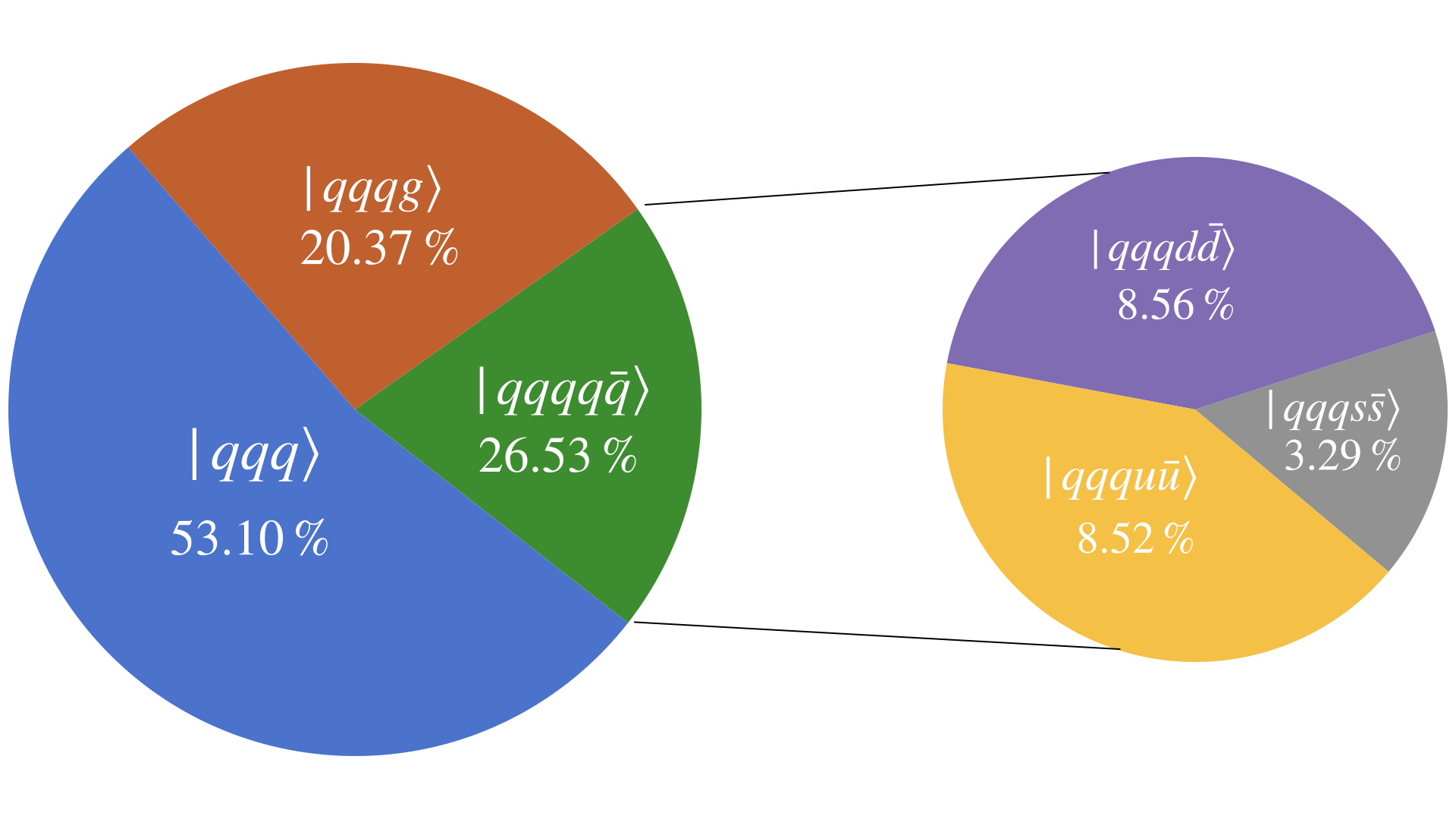}
\caption{Probability distributions of the dominant Fock sectors in the proton, as determined from our BLFQ calculations.}
\label{Fock_probabilities}
\end{center}
\end{figure}

At the model scale, the proton's Fock sector probabilities are distributed as follows: 53.10\% in $|qqq\rangle$, 26.53\% in $|qqqg\rangle$, 8.52\% in $|qqqu\bar{u}\rangle$, 8.56\% in $|qqqd\bar{d}\rangle$, and 3.29\% in $|qqqs\bar{s}\rangle$, as illustrated in Fig.~\ref{Fock_probabilities}. These LFWFs have proven successful in predicting various proton observables, including electromagnetic form factors, charge radii, parton distribution functions, and spin structure~\cite{Xu:2024sjt,Mondal:2025bjn}.

\section{Nucleon Properties within BLFQ}
We generate the nucleon's LFWFs within the BLFQ framework across multiple Fock sector representations by solving for mass eigenstates using either an effective LF Hamiltonian or the LFQCD Hamiltonian. The effective Hamiltonian incorporates key QCD interactions including transverse confinement from LFH, complementary longitudinal confinement, and fixed-coupling OGE, while the QCD Hamiltonian treatment follows the complete formalism appropriate for each Fock sector as discussed in previous sections. These resulting LFWFs enable comprehensive calculations of fundamental nucleon properties spanning electromagnetic and axial FFs, charge radii, PDFs, GPDs, TMDs, angular momentum distributions, gravitational FFs, and double parton distributions. For the present review, we concentrate specifically on three essential aspects of nucleon structure - the electromagnetic FFs revealing the charge and magnetization distributions, the PDFs characterizing the longitudinal momentum structure, and the GPDs providing three-dimensional imaging of the nucleon's quark-gluon structure and their orbital angular momentum.
\subsection{Electromagnetic Form Factors}
In the LF formalism, the Dirac and Pauli FFs of the nucleon, \( F_1(Q^2) \) and \( F_2(Q^2) \), are extracted from the helicity-conserving and helicity-flip matrix elements of the plus component of the electromagnetic current, \( J^+ \equiv \sum_q e_q\, \bar{\psi}_q \gamma^+ \psi_q \), as follows~\cite{Brodsky:1980zm}:
\begin{equation}
\begin{aligned}\label{DFFs}
\langle P+q, \uparrow | \tfrac{J^+(0)}{2P^+} | P, \uparrow \rangle &= F_1(Q^2), \\
\langle P+q, \uparrow | \tfrac{J^+(0)}{2P^+} | P, \downarrow \rangle &= - (q^1 - i q^2)\, \frac{F_2(Q^2)}{2M},
\end{aligned}
\end{equation}
where \( Q^2 = -q^2 \) is the squared momentum transfer, \( M \) is the nucleon mass, and \( e_q \) denotes the electric charge of quark flavor \( q \).

The flavor-separated Dirac \( F_1^q(Q^2) \) and Pauli \( F_2^q(Q^2) \) FFs of the proton can be computed from the LFWFs as overlap integrals~\cite{Brodsky:2000xy}:
\begin{equation}
\begin{aligned}\label{eq_DF}
F_1^q(Q^2) &= \frac{1}{2} \int_{N} \Psi^{N,\,\Lambda\,*}_{\{x_i^\prime,\bm{p}_{i\perp}^{\,\prime},\lambda_i\}} \, \Psi^{N,\,\Lambda}_{\{x_i,\bm{p}_{i\perp},\lambda_i\}}, \\
F_2^q(Q^2) &= -\frac{M}{(q^1-iq^2)} \int_{N} \Psi^{N,\,\Lambda\,*}_{\{x_i^\prime,\bm{p}_{i\perp}^{\,\prime},\lambda_i\}} \, \Psi^{N,\,-\Lambda}_{\{x_i,\bm{p}_{i\perp},\lambda_i\}},
\end{aligned}
\end{equation}
where the integration measure is defined as
\[
\int_{N} \equiv \sum_{N,\,\Lambda,\,\lambda_i} \prod_{i=1}^N \int \left[\frac{{\rm d}x\,{\rm d}^2\bm{p}_{\perp}}{16\pi^3}\right]_i 16\pi^3 \, \delta\left(1-\sum x_j\right) \, \delta^2\left(\sum \bm{p}_{j\perp}\right).
\]

The calculations are performed in a frame where the momentum transfer is purely transverse, \( q = (0,\, 0,\, \bm{q}_{\perp}) \), so that \( Q^2 = -q^2 = \bm{q}_{\perp}^{\,2} \). In this frame, the longitudinal momentum fractions remain unchanged, \( x_i^\prime = x_i \), while the transverse momenta transform differently for struck and spectator partons:
\begin{itemize}
    \item Struck quark: \(\bm{p}_{1\perp}^{\,\prime} = \bm{p}_{1\perp} + (1 - x_1) \bm{q}_{\perp}\),
    \item Spectator partons: \(\bm{p}_{i\perp}^{\,\prime} = \bm{p}_{i\perp} - x_i \bm{q}_{\perp}\).
\end{itemize}

The full nucleon FFs are reconstructed by combining the flavor-separated contributions~\cite{Cates:2011pz,Chakrabarti:2013dda,Mondal:2016xpk}. The Sachs electric and magnetic FFs, \( G_{\rm E}(Q^2) \) and \( G_{\rm M}(Q^2) \), are defined in terms of the Dirac and Pauli FFs as:
\begin{equation}
\begin{aligned}\label{eq_Sachs}
G_{\rm E}(Q^2) &= F_1(Q^2) - \frac{Q^2}{4M^2} F_2(Q^2), \\
G_{\rm M}(Q^2) &= F_1(Q^2) + F_2(Q^2).
\end{aligned}
\end{equation}
These relations connect the theoretical form factors \( F_1 \) and \( F_2 \) to the experimentally measurable electric and magnetic distributions of the nucleon.

\begin{figure}
\begin{center}
\includegraphics[width=0.5\linewidth]{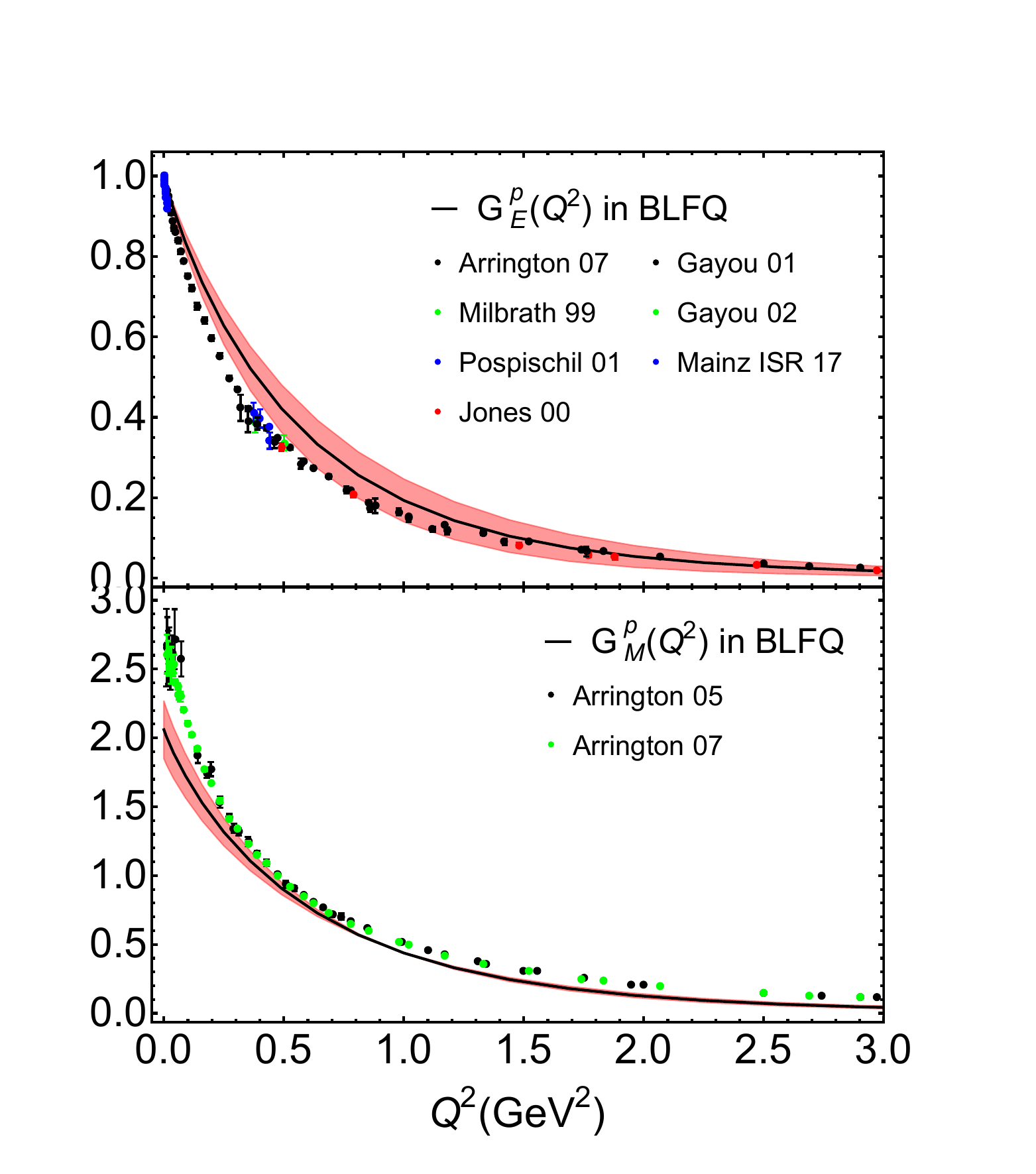}
\caption{Proton electromagnetic FFs computed using LFWFs from the leading three Fock sectors: \( |qqq\rangle \), \( |qqqg\rangle \), and \( |qqqq\bar{q}\rangle \) as reported in Ref.~\cite{Xu:2024sjt}. The black lines show our BLFQ predictions, with red bands reflecting uncertainties from model parameters. Experimental data for the electric form factor \( G^{p}_{\rm E} \) are taken from Refs.~\cite{Gayou:2001qt,JeffersonLabHallA:1999epl,Arrington:2007ux,JeffersonLabHallA:2001qqe,A1:2001xxy,BatesFPP:1997rpw}, and for the magnetic form factor \( G^{p}_{\rm M} \) from Refs.~\cite{Arrington:2004ae,Arrington:2007ux}.
}
\label{proton_FFs}
\end{center}
\end{figure}
\begin{figure}
\begin{center}
\includegraphics[width=0.45\linewidth]{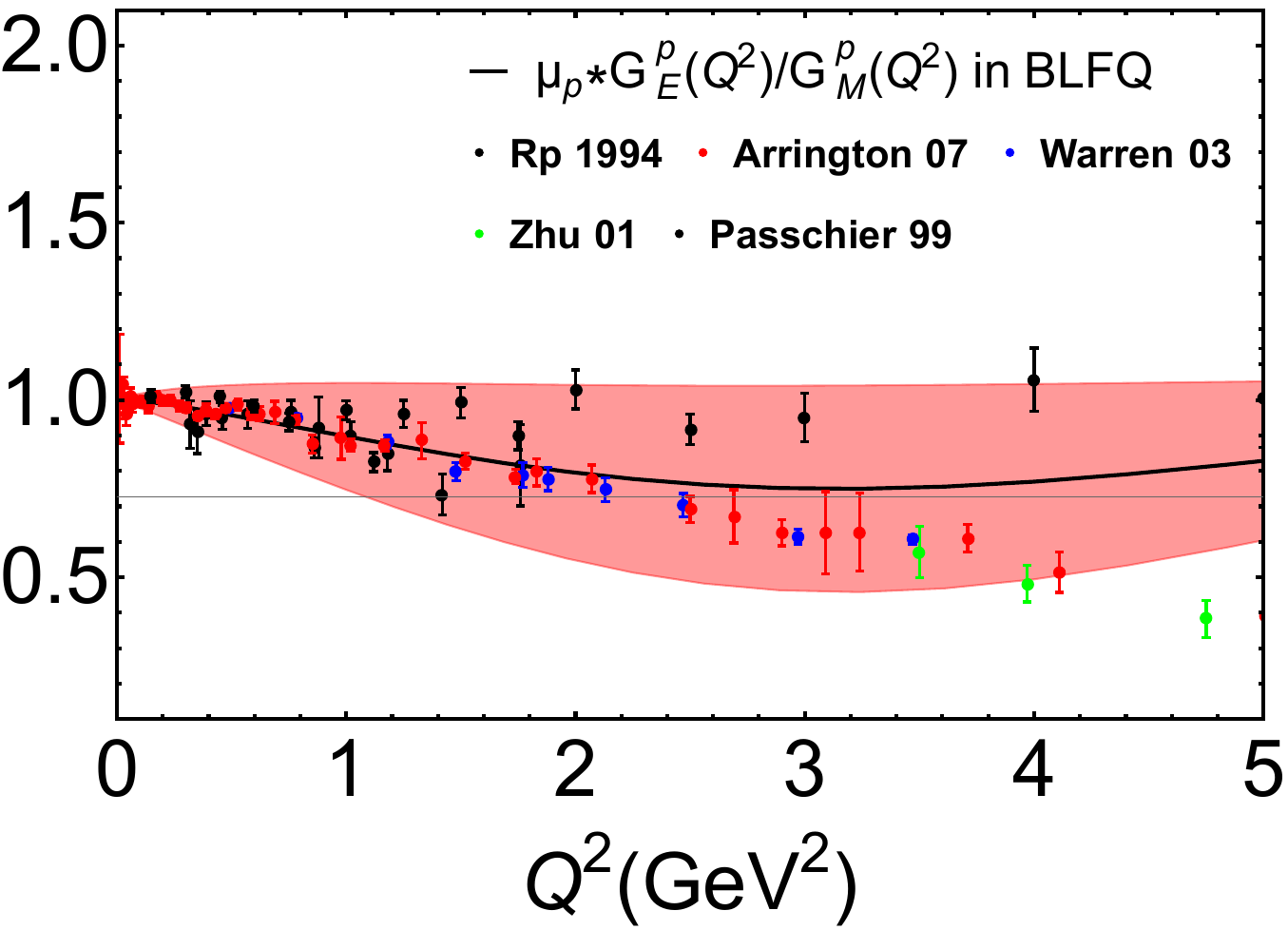}
\caption{Ratio of the proton’s Sachs FFs, $R_{p} = \mu_{p} G_{\rm E}^{p} / G_{\rm M}^{p}$. The black line shows the BLFQ prediction adopted from Ref.~\cite{Xu:2024sjt}, with the red band indicating uncertainties from the model parameters. Experimental data are taken from Refs.~\cite{JeffersonLabHallA:2001qqe,A1:2001xxy,BatesFPP:1997rpw,Punjabi:2005wq,JeffersonLabHallA:2011yyi,Walker:1993vj,Zhan:2011ji,MacLachlan:2006vw,Paolone:2010qc}.
}
\label{proton_FFs_ratio}
\end{center}
\end{figure}

The proton electromagnetic FFs calculated using the LFWFs in the leading three Fock sectors—namely, $|qqq\rangle$, $|qqqg\rangle$, and $|qqqq\bar{q}\rangle$—are shown in Fig.~\ref{proton_FFs} following results in Ref.~\cite{Xu:2024sjt}. The red bands represent theoretical uncertainties arising from variations in the model parameters discussed earlier. Overall, our results exhibit good agreement with the experimental data for the proton electric FF. For the magnetic FF, the agreement is satisfactory at large-$Q^2$, although a deviation of approximately 25\% is observed in the low-$Q^2$ region.

In Fig.~\ref{proton_FFs_ratio}, we present a comparison between our prediction for the ratio of Sachs FFs, $R_{p} = \mu_{p} G_{\rm E}^{p} / G_{\rm M}^{p}$, and available experimental data. Within our approach, relatively large constituent quark masses are employed to effectively emulate the effects of higher Fock sectors. This modeling choice renders the quarks more nonrelativistic at long distances (low-$Q^2$) and more relativistic at short distances (high-$Q^2$). Consequently, the $Q^2$ dependence of the $G_{\rm E}/G_{\rm M}$ ratio is accurately reproduced in the low-$Q^2$ domain, while deviations emerge at higher $Q^2$, reflecting the increasing importance of relativistic effects.

Experimental measurements of the proton Dirac FF $F_1^p$~\cite{Arnold:1986nq,Sill:1992qw} have shown reasonable agreement with the pQCD scaling prediction, $F_1^p \propto 1/Q^4$~\cite{Lepage:1979za}. Nevertheless, it has been argued that pQCD may not be fully applicable to exclusive processes at the momentum transfers currently accessible in experiments~\cite{Isgur:1989cy}. In particular, Jefferson Lab measurements of the ratio $F_2^p/F_1^p$~\cite{Gayou:2001qt,JeffersonLabHallA:1999epl,JeffersonLabHallA:2001qqe,Punjabi:2005wq,Puckett:2010ac} show significant deviations from the original pQCD prediction, $F_2^p/F_1^p \propto 1/Q^2$~\cite{Lepage:1979za,Chakrabarti:2013dda,Ahmady:2021qed}.

\begin{figure}
\begin{center}
\includegraphics[width=0.5\linewidth]{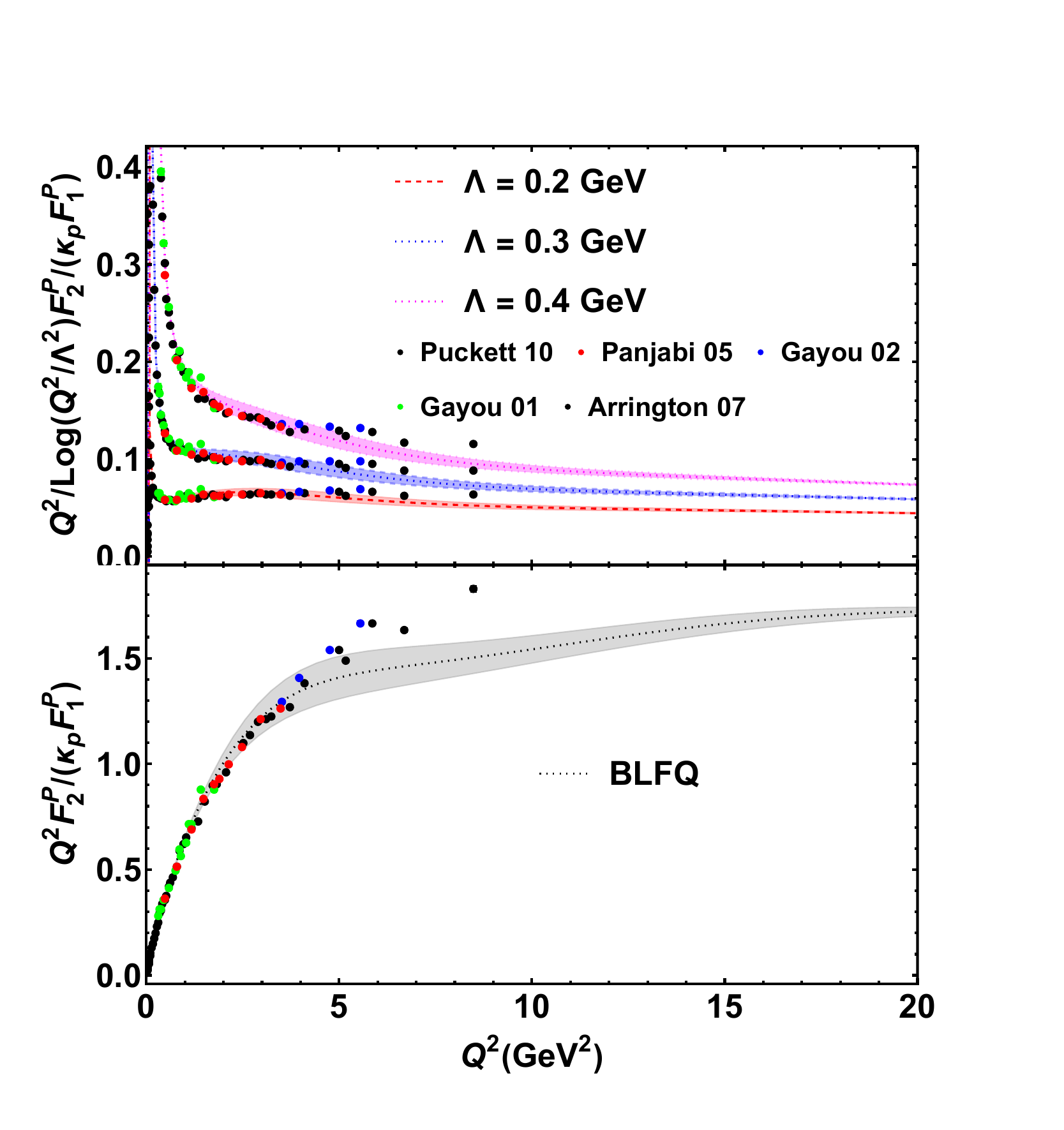}
\caption{Ratio of the proton Pauli to Dirac FFs. The lower panel shows the quantity \( Q^2 F_2^p / (\kappa_p F_1^p) \), while the upper panel presents the scaled ratio \( \left(Q^2 / \log^2[Q^2 / \Lambda_{\rm QCD}^2]\right) F_2^p / (\kappa_p F_1^p) \) for three values of the QCD scale parameter: \( \Lambda_{\rm QCD} = 200 \), 300, and 400 MeV. Here, \( \kappa_p \) denotes the anomalous magnetic moment of the proton. Experimental data are taken from Refs.~\cite{Gayou:2001qt,JeffersonLabHallA:1999epl,JeffersonLabHallA:2001qqe,Punjabi:2005wq,Puckett:2010ac}.
}
\label{F1F2ratio}
\end{center}
\end{figure}

Instead, the data are found to align more closely with a modified pQCD prediction: $Q^2 F_2^p/F_1^p \propto \log^2(Q^2/\Lambda_{\rm QCD}^2)$~\cite{Belitsky:2002kj,Brodsky:2003pw}, where $\Lambda_{\rm QCD}$ is the QCD scale parameter. This behavior appears to hold even at relatively large $Q^2$. Our results for the scaling of the $F_2^p/F_1^p$ ratio, shown in Fig.~\ref{F1F2ratio}, are in reasonable agreement with this revised pQCD behavior.

The electromagnetic radii of the proton are extracted from the slopes of the Sachs FFs at $Q^2 = 0$~\cite{Ernst:1960zza}. We obtain a charge radius of $\sqrt{\langle r_{\rm E}^2 \rangle} = 0.72 \pm 0.05$~fm and a magnetic radius of $\sqrt{\langle r_{\rm M}^2 \rangle} = 0.73 \pm 0.02$~fm. These results are to be compared with the experimental determinations: $\sqrt{\langle r_{\rm E}^2 \rangle}_{\rm exp} = 0.840^{+0.003}_{-0.002}$~fm and $\sqrt{\langle r_{\rm M}^2 \rangle}_{\rm exp} = 0.849 \pm 0.003$ fm~\cite{Lin:2021xrc,ParticleDataGroup:2024cfk}.

\subsection{Axial Form Factor}
An essential quantity for understanding nucleon structure is the isovector axial FF of the nucleon. At zero momentum transfer, this FF defines the nucleon axial charge, \( g_A \). The \( Q^2 \)-dependence of the axial FF plays a crucial role in interpreting experimental processes such as elastic neutrino-nucleon scattering. Accurate modeling of such processes is vital for achieving the precision required in determining neutrino oscillation parameters~\cite{NuSTEC:2017hzk}.

The isovector axial current of the light quarks in QCD is given by
\begin{align}
A^\mu_a(z) = \bar{\psi}_q(z)\gamma^{\mu}\gamma^5\frac{\tau^a}{2}\psi_q(z),
\end{align}
where \( \tau^a \) (\( a = 1, 2, 3 \)) are the Pauli isospin matrices. In the isospin symmetry limit, the matrix element of this current between nucleon states is parametrized as
\begin{align}
\braket{P+q, \Lambda'|A^\mu_a(0)|P, \Lambda}
= \bar{u}(P+q, \Lambda')\left[\gamma^{\mu}G_{\rm A}(Q^2) + \frac{q^{\mu}}{2M}G_{\rm P}(Q^2)\right]\gamma^5 \frac{\tau^a}{2} u(P,\Lambda),
\end{align}
where \( G_{\rm A}(Q^2) \) and \( G_{\rm P}(Q^2) \) denote the axial and induced pseudoscalar form factors, respectively.

In the LF formalism, the axial FF can be expressed—analogous to the electromagnetic case—in terms of overlaps of LFWFs using the plus component of the axial current. Specifically,
\begin{align}
G^q_{\rm A}(Q^2) = \frac{1}{2} \int_{{N}} \lambda_1 \,\Psi^{{N},\,\Lambda\,*}_{\{x_i^\prime,\bm{p}_{i\perp}^{\,\prime},\lambda_i\}} \, \Psi^{{N},\,\Lambda}_{\{x_i,\bm{p}_{i\perp},\lambda_i\}},
\end{align}
where \( \lambda_1 = 1\,(-1) \) for a struck quark with positive (negative) helicity. The quantity \( G^q_{\rm A}(Q^2) \) denotes the flavor axial FF.

\begin{figure}[htbp]
\centering
\includegraphics[width=0.5\columnwidth]{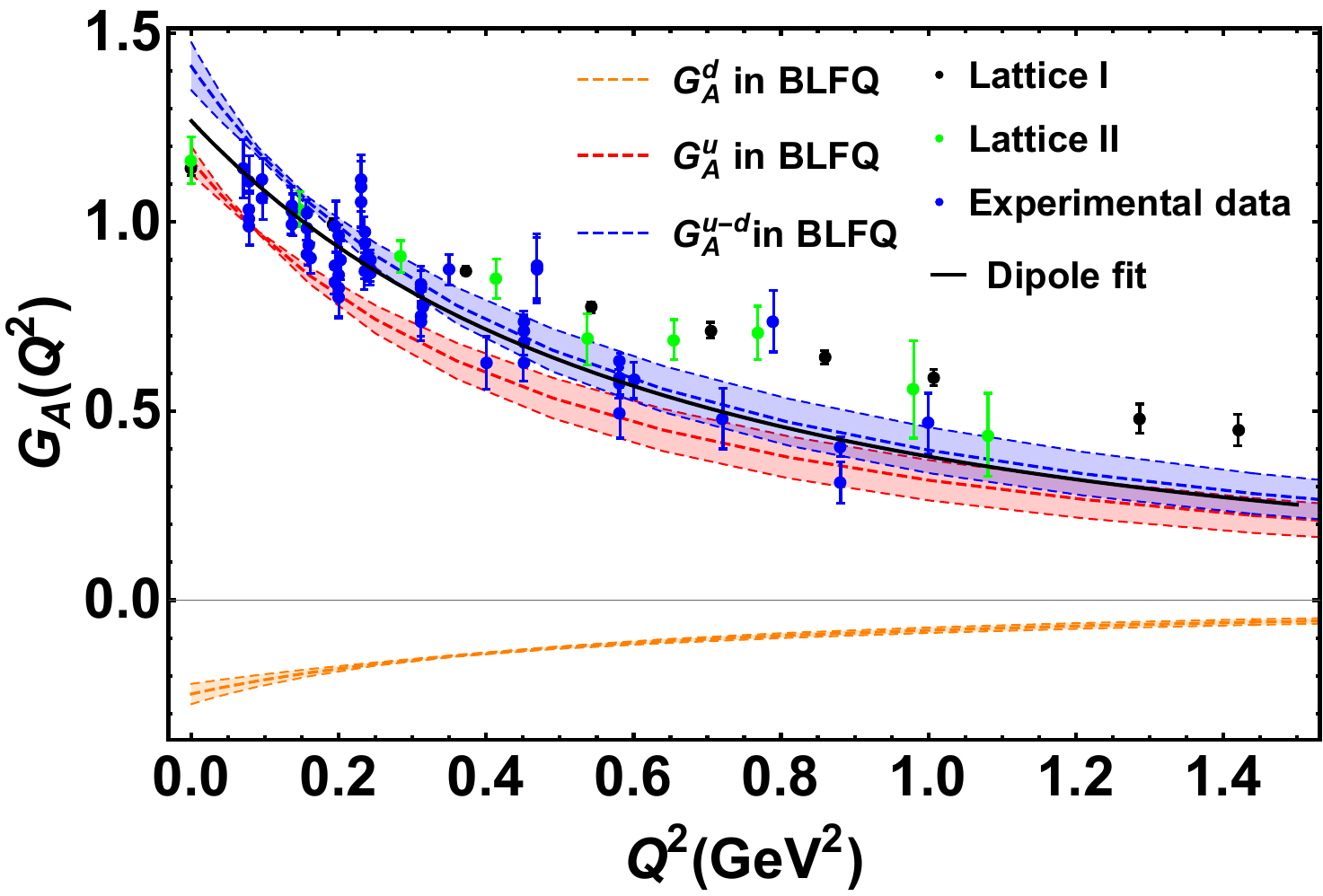}
\caption{Axial FFs \( G_{\rm A} = G_{\rm A}^u - G_{\rm A}^d \), along with the individual contributions \( G_{\rm A}^u \) and \( G_{\rm A}^d \), as functions of \( Q^2 \), obtained using nucleon LFWFs from results obtained using only the leading Fock-sector component~\cite{Xu:2021wwj}. The BLFQ results are shown as the blue band for \( G_{\rm A} \), the pink band for \( G_{\rm A}^u \), and the orange band for \( G_{\rm A}^d \). Experimental data are taken from Refs.~\cite{Bernard:2001rs,Hashamipour:2019pgy}, and lattice QCD results from Ref.~\cite{Alexandrou:2013joa}. The black line represents the dipole fit to experimental data~\cite{Bernard:2001rs}.
}
\label{AFF}
\end{figure}
Figure~\ref{AFF} presents our results for the nucleon isovector axial form factor, \( G_{\rm A}(Q^2) = G_{\rm A}^u(Q^2) - G_{\rm A}^d(Q^2) \), along with the individual contributions from the up and down quarks. These results are obtained using the nucleon LFWFs corresponding to the leading Fock-sector component, derived from mass eigenstates of a LF effective Hamiltonian. This Hamiltonian includes a transverse confining potential inspired by LFH, longitudinal confinement, and a OGE interaction with fixed coupling~\cite{Mondal:2019jdg,Xu:2021wwj}, see Eq.~\eqref{eq:hamiltonian}.

Our predictions are compared with experimental data from (anti)neutrino scattering on protons or nuclei, as well as from charged pion electroproduction~\cite{Bernard:2001rs,Schindler:2006jq,Hashamipour:2019pgy}, and with lattice QCD results from Ref.~\cite{Alexandrou:2013joa}. Taking into account both the experimental uncertainties and those associated with this implementation of the BLFQ framework, we find good overall agreement with the data.

For reference, experimental measurements of the axial FF are well described by the dipole parametrization:
\begin{align}
G_{\rm A}(Q^2) = \frac{g_{\rm A}}{(1+Q^2/M_{\rm A}^2)^2},\label{dipole_fit}
\end{align}
with \( g_{\rm A} = 1.2673 \pm 0.0035 \) and axial mass \( M_{\rm A} = 1.1~\text{GeV} \)~\cite{Bernard:2001rs}. Our results are consistent with this dipole form.

At zero momentum transfer, \( G_{\rm A}(0) \) defines the axial charge, \( g_{\rm A} \), which is precisely determined from neutron beta decay: \( g_{\rm A} = 1.2723 \pm 0.0023 \)~\cite{ParticleDataGroup:2018ovx}. Our calculation yields \( g_{\rm A} = 1.41 \pm 0.06 \), which is somewhat larger than both the experimental and lattice QCD values~\cite{Alexandrou:2017oeh,Yao:2017fym}. This axial charge reflects the difference in spin contributions from up and down quarks in the proton. Specifically, our model slightly overestimates the up-quark contribution and underestimates the down-quark contribution relative to experiment and lattice results.

Summing over flavors, we compute the total quark spin contribution to the proton spin. Within our model, which includes only the leading Fock sector, quark spin accounts for approximately 91\% of the proton spin. This is notably higher than the value inferred from experiment, where quark spin contributes only about 40\%~\cite{Leader:2010rb}. This significant discrepancy highlights the importance of including higher Fock sectors in the model. The presence of dynamical gluons and sea quarks is expected to reduce the quark spin contribution and enhance the role of quark and gluon orbital angular momentum in the proton spin decomposition.

Finally, we extract the proton axial radius from the slope of \( G_{\rm A}(Q^2) \) at \( Q^2 = 0 \). Our result, \( \sqrt{\langle r_{\rm A}^2 \rangle} = 0.680 \pm 0.072~\text{fm} \), agrees very well with the value obtained from analyses of neutrino-nucleon scattering data, \( \sqrt{\langle r_{\rm A}^2 \rangle} = 0.667 \pm 0.120~\text{fm} \)~\cite{Hill:2017wgb,Meyer:2016oeg}.
\subsection{Parton Distribution Functions}
PDFs characterize the probability of finding a parton carrying a specific fraction of the longitudinal momentum of a hadron in LF dynamics. They offer valuable insights into the nonperturbative internal structure of hadrons. The quark PDF of the nucleon, which encodes the longitudinal momentum and spin distributions of quarks, is defined as~\cite{Meissner:2007rx}
\begin{align}\label{defi_pdf}
\Phi^{\Gamma(q)}(x)= \frac{1}{2}\int \frac{d z^-}{2\pi} e^{i x P^+ z^-} 
\left\langle P,\Lambda\left|\bar{\psi}_{q}\left(-\tfrac{1}{2}z\right)\Gamma\,\psi_{q}\left(\tfrac{1}{2}z\right)\right|P, \Lambda\right\rangle 
\bigg|_{z^+=\vec{z}_\perp=0},
\end{align}
where $\Gamma$ specifies the Dirac matrix structure determining the type of distribution. For example, $\Gamma = \gamma^+$ gives the unpolarized PDF $f(x)$, $\Gamma = \gamma^+ \gamma^5$ yields the helicity PDF $g_1(x)$, and $\Gamma = i \sigma^{j+} \gamma^5$ corresponds to the transversity PDF $h_1(x)$. We work in the LF gauge, $A^+ = 0$, in which the gauge link between the quark fields reduces to unity.

The unpolarized gluon PDF is analogously defined as~\cite{Meissner:2007rx}
\begin{align}\label{defi_pdf_g}
\Phi^{(g)}(x) = \frac{1}{xP^+} \int \frac{d z^-}{2\pi} e^{i x P^+ z^-} 
\left\langle P,\Lambda\left|G^{+\mu}\left(-\tfrac{1}{2}z\right)G^{+}_{\mu}\left(\tfrac{1}{2}z\right)\right|P, \Lambda\right\rangle 
\bigg|_{z^+=\vec{z}_\perp=0}.
\end{align}
For the gluon helicity distribution, the field strength tensor $G^+_\mu$ is replaced by its dual $\tilde{G}^+_\mu$, with $\tilde{G}^{\alpha\beta} = \frac{1}{2} \epsilon^{\alpha\beta\gamma\delta} G_{\gamma\delta}$. Notably, a gluon transversity PDF does not exist for spin-$\tfrac{1}{2}$ hadrons, as helicity-flip transitions for gluons require spin-1 or higher hadronic states to conserve angular momentum~\cite{Kumano:2019igu}.

Using the LFWFs, the unpolarized, helicity, and transversity quark PDFs can be expressed as:
\begin{equation}
\begin{aligned}
f(x) &= \int_{{N}} \frac{1}{2} \, \Psi^{{N},\,\Lambda\,*}_{\{x_i,\bm{p}_{i\perp},\lambda_i\}} \, 
\Psi^{N,\,\Lambda}_{\{x_i,\bm{p}_{i\perp},\lambda_i\}} \, \delta(x - x_i), \\
\Delta f(x) &= \int_{N} \frac{\lambda_1}{2} \, \Psi^{N,\,\Lambda\,*}_{\{x_i,\bm{p}_{i\perp},\lambda_i\}} \, 
\Psi^{N,\,\Lambda}_{\{x_i,\bm{p}_{i\perp},\lambda_i\}} \, \delta(x - x_i), \\
\delta f(x) &= \int_{N} \Psi^{N,\,\Lambda\,*}_{\{x_i,\bm{p}_{i\perp},\lambda_i^\prime\}} \, 
\Psi^{N,\,-\Lambda}_{\{x_i,\bm{p}_{i\perp},\lambda_i\}} \, \delta(x - x_i),
\label{eqn:pdf_i}
\end{aligned}
\end{equation}
where the index $N$ denotes different Fock sectors. 
In the transversity distribution, the struck parton flips its helicity, i.e., $\lambda_1' = -\lambda_1$, while spectator helicities remain unchanged ($\lambda_i' = \lambda_i$ for $i \neq 1$).

Since PDFs are interpreted as number densities, they satisfy the normalization condition for valence quarks:
\begin{align}
    \int_{0}^1 f(x) \, \mathrm{d}x = n_q,
\end{align}
where $n_q = 2$ for up quarks and $n_q = 1$ for down quarks in the proton. Including gluon and sea quark contributions, the momentum sum rule ensures:
\begin{align}
    \int_0^1 \sum_{i} x f^i(x) \, \mathrm{d}x = 1.
\end{align}

To evolve our PDFs from the low model scale to higher energy scales, we employ the \texttt{HOPPET} toolkit~\cite{Salam:2008qg}, which numerically solves the QCD Dokshitzer–Gribov–Lipatov–Altarelli–Parisi (DGLAP) equations~\cite{Dokshitzer:1977sg,Gribov:1972ri,Altarelli:1977zs}. The initial scales of our models are fixed by matching the second moment of the total valence quark distribution at $\mu^2 = 10~\text{GeV}^2$ to the global fit result, yielding $\langle x \rangle_{u+d} = 0.37 \pm 0.01$~\cite{NNPDF:2017mvq,Cocuzza:2021cbi}.

\begin{figure}
\centering
\includegraphics[width=.5\columnwidth]{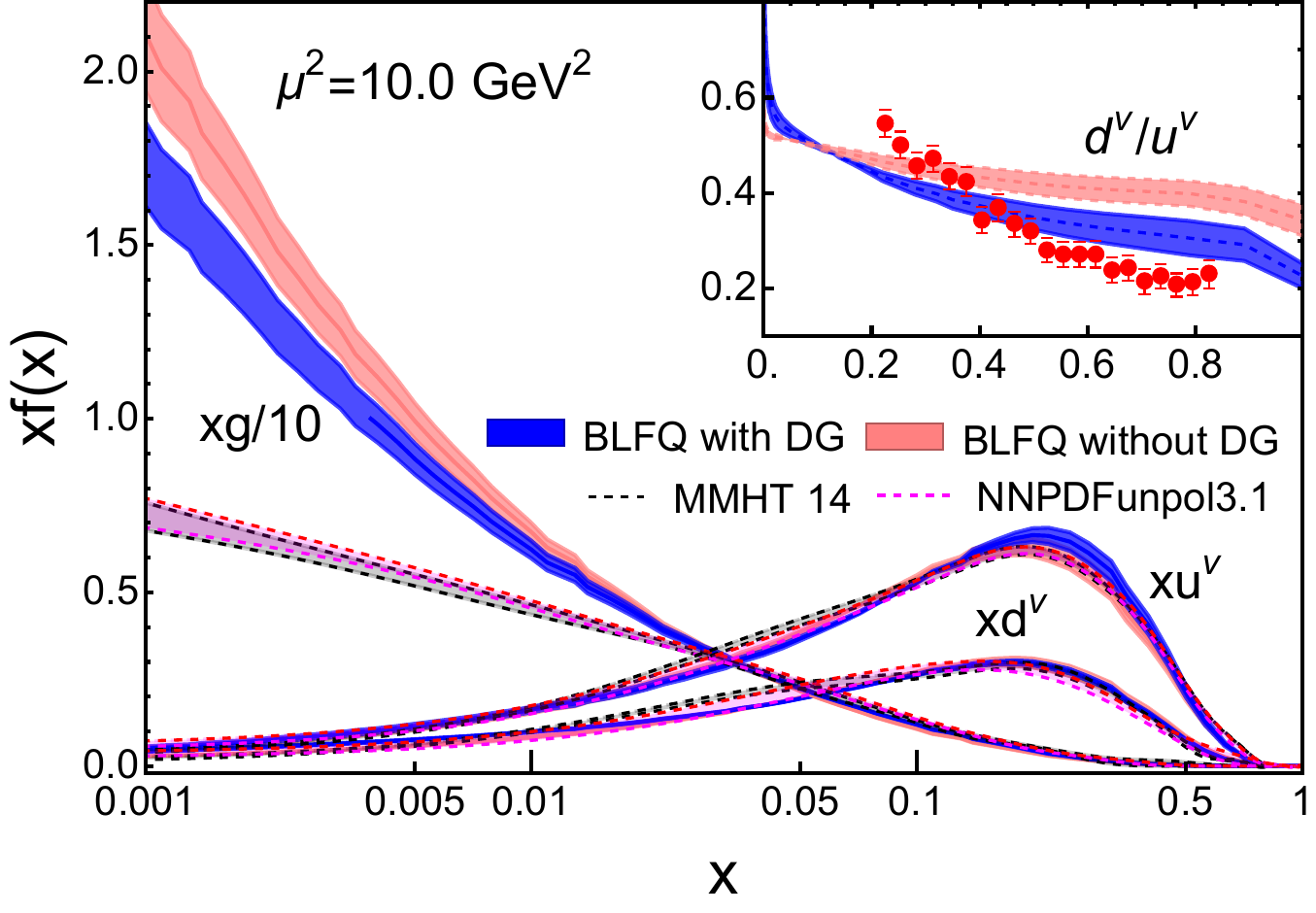}
\caption{Unpolarized valence quark and gluon PDFs of the proton computed using LFWFs from the leading two Fock sectors: \( |qqq\rangle \) and \( |qqqg\rangle \) as reported in Ref.~\cite{Xu:2022yxb}. The BLFQ results are shown as blue bands (including one dynamical gluon ``DG") and pink bands (based on a LF effective Hamiltonian with only the valence Fock component~\cite{Xu:2021wwj}). These results are compared with the global fits from NNPDF3.1~\cite{NNPDF:2017mvq} and MMHT~\cite{Harland-Lang:2014zoa}. \textbf{Inset:} The ratio of valence quark PDFs is compared with the extracted data from the JLab MARATHON experiment~\cite{JeffersonLabHallATritium:2021usd}.}
\label{unpolarized_pdf}
\end{figure}
\begin{figure}[htp]
\begin{center}
\includegraphics[width=0.5\linewidth]{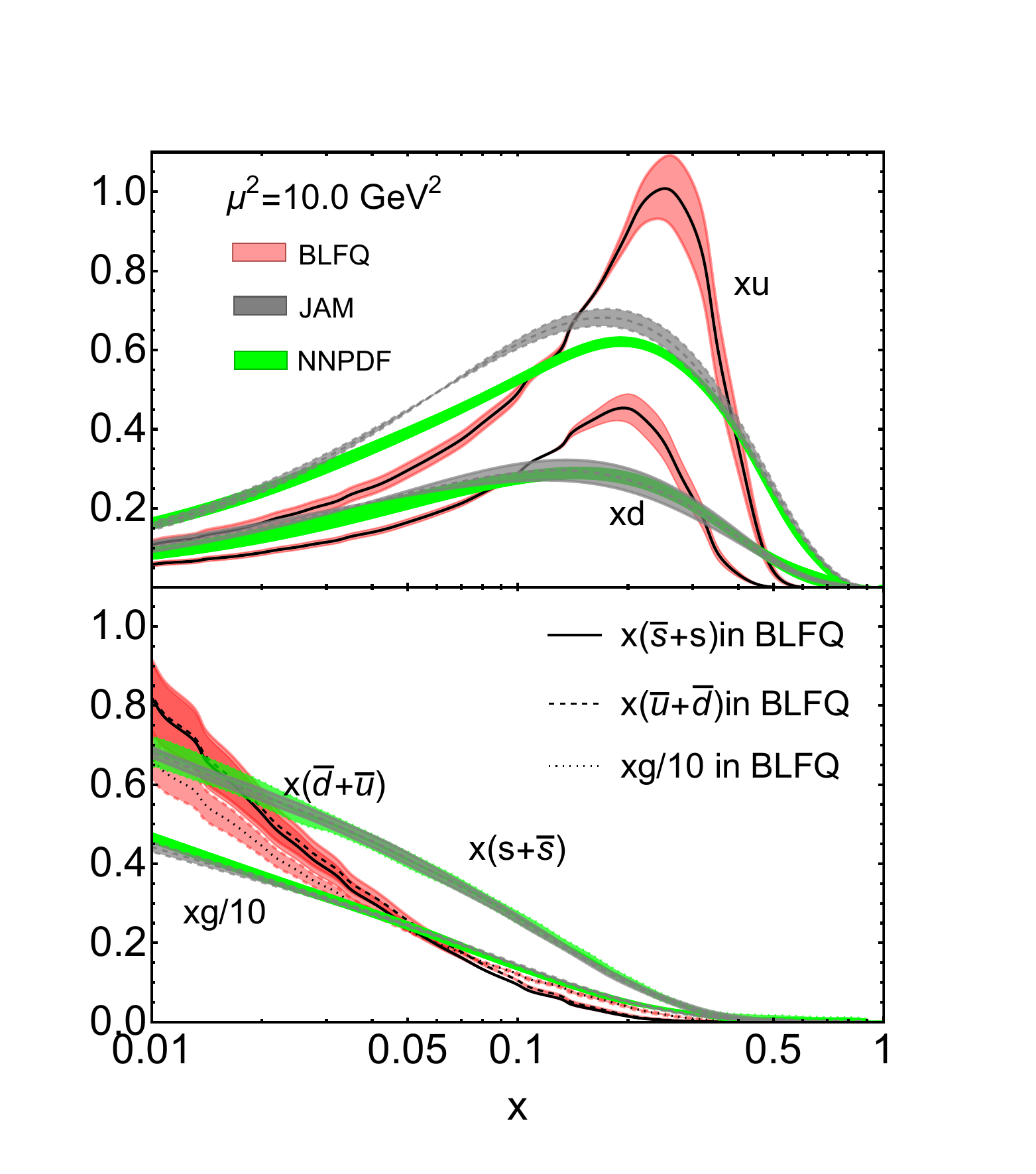}
\caption{Proton unpolarized PDFs calculated using LFWFs including the leading three Fock sectors: \( |qqq\rangle \), \( |qqqg\rangle \), and \( |qqqq\bar{q}\rangle \) as reported in Ref.~\cite{Xu:2024sjt}. Our results are shown as red bands, consistent with previous figures, and are compared with global QCD analyses from NNPDF3.1~\cite{NNPDF:2017mvq} (light green bands) and JAM~\cite{Cocuzza:2021cbi} (gray bands).}
\label{fig_pdf_5quarks}
\end{center}
\end{figure}
\subsubsection{Unpolarized PDFs}
Figure~\ref{unpolarized_pdf} presents our results for the proton unpolarized PDFs at the scale $\mu^2 = 10~\mathrm{GeV}^2$. We compare our valence quark and gluon distributions, obtained after QCD evolution, with global fits from NNPDF3.1~\cite{NNPDF:2017mvq} and MMHT~\cite{Harland-Lang:2014zoa}. 

In this study, our approach includes both the leading Fock sector with three constituent quarks and an additional sector containing three quarks and a dynamical gluon. For comparison, we also show the proton PDFs obtained from a LF effective Hamiltonian based on a valence Fock representation~\cite{Mondal:2019jdg,Xu:2021wwj}. The error bands in our evolved distributions reflect an assumed $10\%$ uncertainty in the initial model scale.

We find good agreement between our predictions for the proton's valence quark PDFs and the global fits. In particular, the ratio $d^v(x)/u^v(x)$ is consistent with the extraction from the MARATHON experiment at Jefferson Lab~\cite{JeffersonLabHallATritium:2021usd}. A recent analysis and extrapolation of the MARATHON data yields $\lim_{x \to 1} d^v/u^v = 0.230 \pm 0.057$~\cite{Cui:2021gzg}, while our prediction gives $d^v/u^v = 0.225 \pm 0.025$ as $x \to 1$, in excellent agreement.

According to the Drell-Yan–West relation~\cite{Drell:1969km,West:1970av}, valence quark distributions at large $\mu^2$ are expected to fall off as $(1-x)^p$, where $p$ depends on the number of valence constituents, with $p = 3$ for the proton. We find that the up-quark PDF behaves as $(1-x)^{3.2 \pm 0.1}$, while the down-quark PDF falls off as $(1-x)^{3.5 \pm 0.1}$, in line with perturbative QCD predictions~\cite{Brodsky:1994kg}.

Our results also indicate that the gluon PDF improves at small $x$ and approaches the global fits~\cite{NNPDF:2017mvq,Harland-Lang:2014zoa} upon including the dynamical gluon component. For $x \gtrsim 0.05$, the gluon distribution shows reasonable agreement with the global fits.

Figure~\ref{fig_pdf_5quarks} shows the evolved unpolarized PDFs of the proton at $\mu^2 = 10~\text{GeV}^2$, obtained from a LFQCD Hamiltonian that includes the three-quark, three-quark–gluon, and three-quark–quark–antiquark Fock sectors, without introducing an explicit confining potential. We compare our results for the valence quark and gluon distributions with global analyses from NNPDF3.1~\cite{NNPDF:2017mvq} and JAM~\cite{Cocuzza:2021cbi}. 

Our valence quark distributions exhibit qualitative agreement with the global fits. However, the relatively large constituent quark masses used in our framework, together with the absence of a longitudinal confining potential, limit the modeling of longitudinal excitations. As a result, the valence PDFs appear narrower than those from global fits, particularly for $x > 0.1$.

The gluon PDF shows good agreement with global analyses in the intermediate and large-$x$ region ($x > 0.05$). At small $x$, where DGLAP evolution is especially sensitive to the choice of the initial model scale $\mu_0$, our gluon distribution still maintains qualitative consistency with the global fits.

\begin{figure}
\centering
\includegraphics[width=.45\columnwidth]{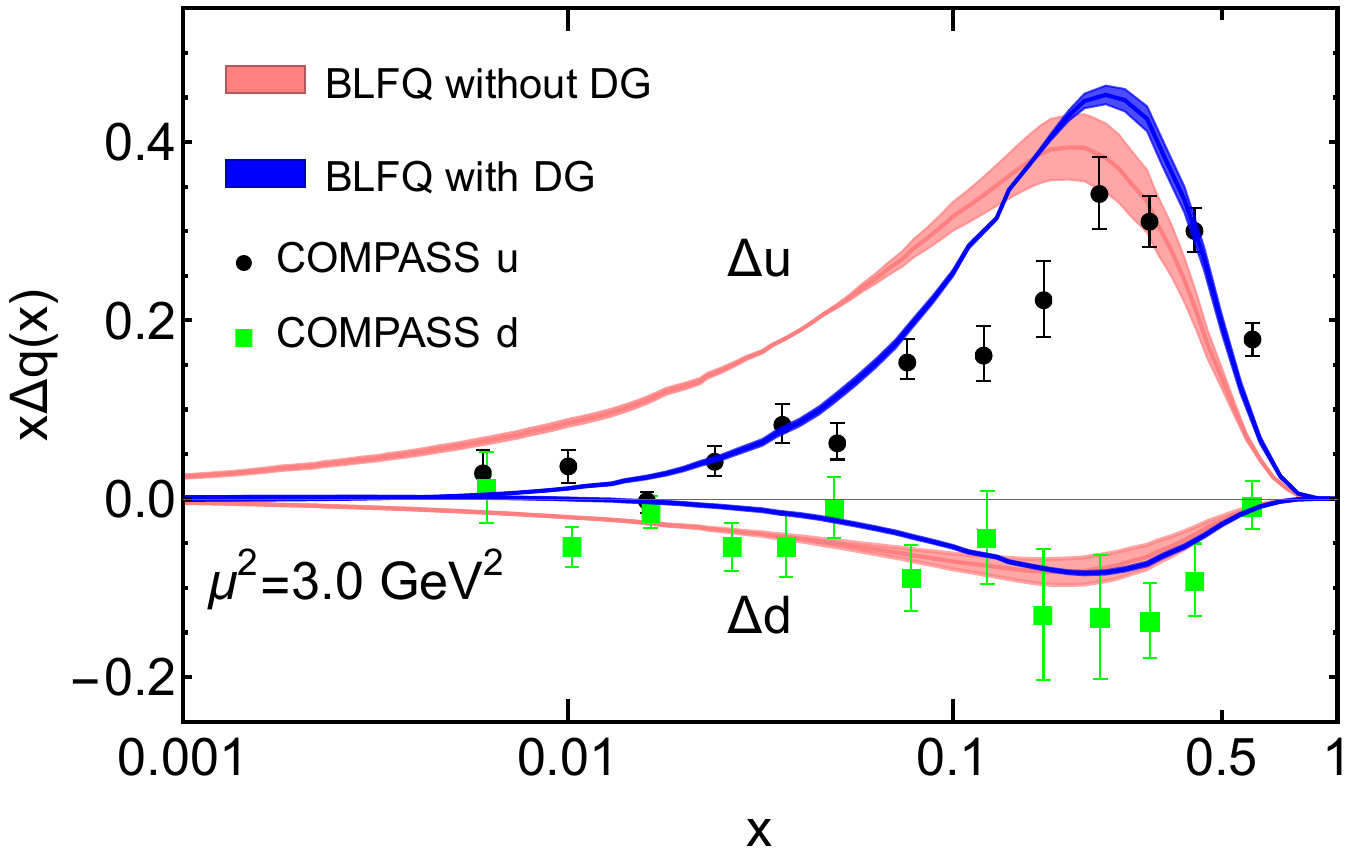}
\includegraphics[width=.45\columnwidth]{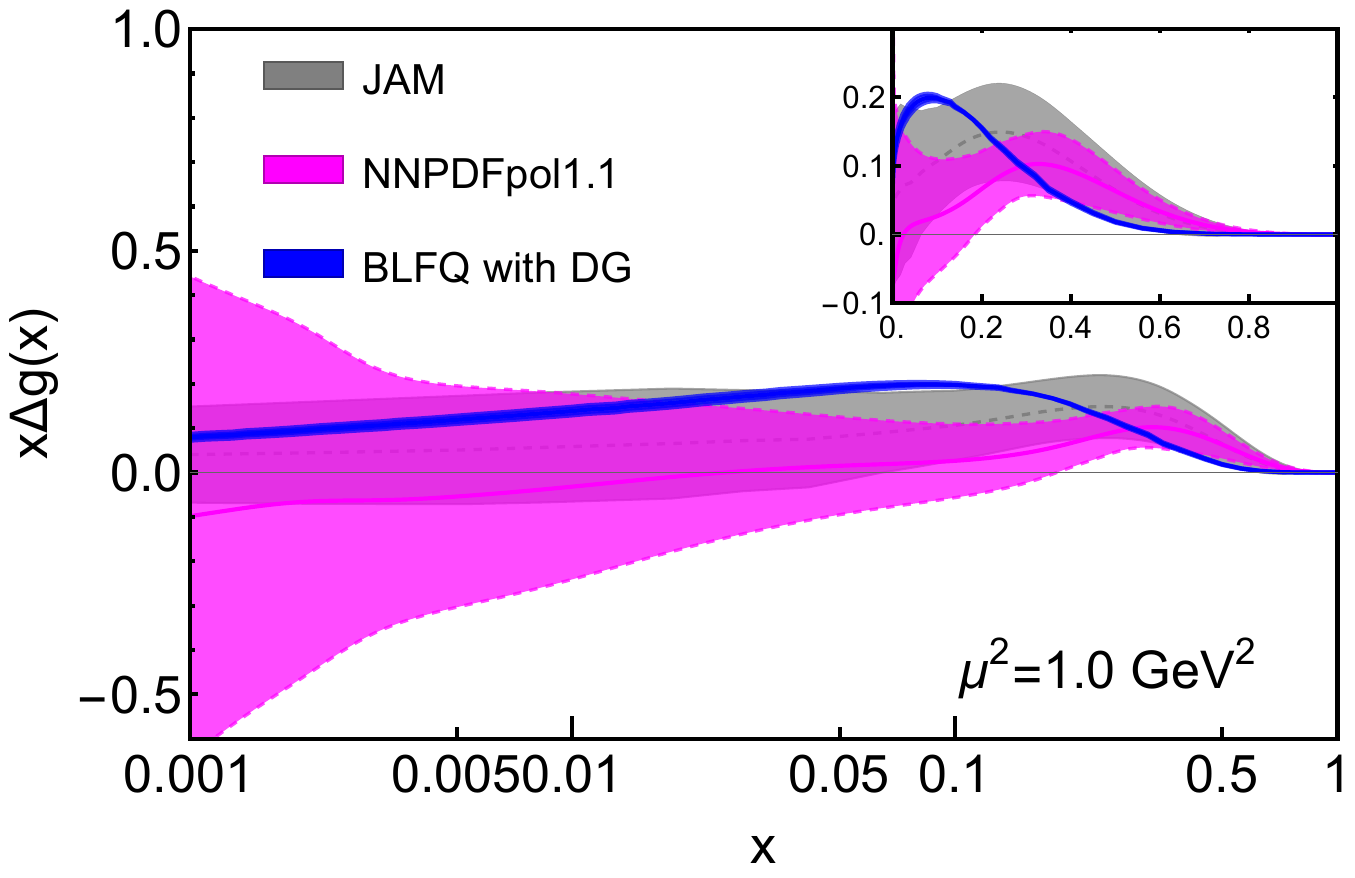}
\caption{Left panel: Helicity PDFs for valence quarks in the proton. BLFQ predictions are shown as blue bands (including one dynamical gluon ``DG"~\cite{Xu:2022yxb}) and pink bands (from a LF effective Hamiltonian with only the valence Fock sector~\cite{Xu:2021wwj}), compared to data from the COMPASS Collaboration~\cite{COMPASS:2010hwr}. Right panel: Gluon helicity PDF in the proton. The BLFQ result (blue bands) is compared with global analyses by JAM (gray band)~\cite{Sato:2016tuz} and NNPDFpol1.1 (magenta band)~\cite{Nocera:2014gqa}. The inset displays the gluon helicity PDF on a linear scale.
}
\label{helicity_pdf_12}
\end{figure}
\subsubsection{Helicity PDFs}
The helicity PDFs obtained from effective Hamiltonians including the valence Fock sector ($|qqq\rangle$) alone and the valence sector plus an additional three-quark–one-gluon Fock sector ($|qqq\rangle$ and $|qqqg\rangle$) are shown in Fig.~\ref{helicity_pdf_12} at the scale $\mu^2 = 3$ GeV$^2$, for the up and down quarks in the proton. Our BLFQ predictions are compared with experimental data from COMPASS~\cite{COMPASS:2010hwr}.

We find that the down quark helicity PDF is in reasonable agreement with the COMPASS measurements. For the up quark, the helicity distribution $x\Delta f(x)$ computed in the valence-only sector tends to be overestimated at low $x$, but approaches the data well for $x \gtrsim 0.25$. We notice that $\Delta u(x)$ improves significantly at small-$x$ region with two Fock-sectors treatment for the nucleon with dynamical gluon.

\begin{figure}
\begin{center}
\includegraphics[width=0.6\linewidth]{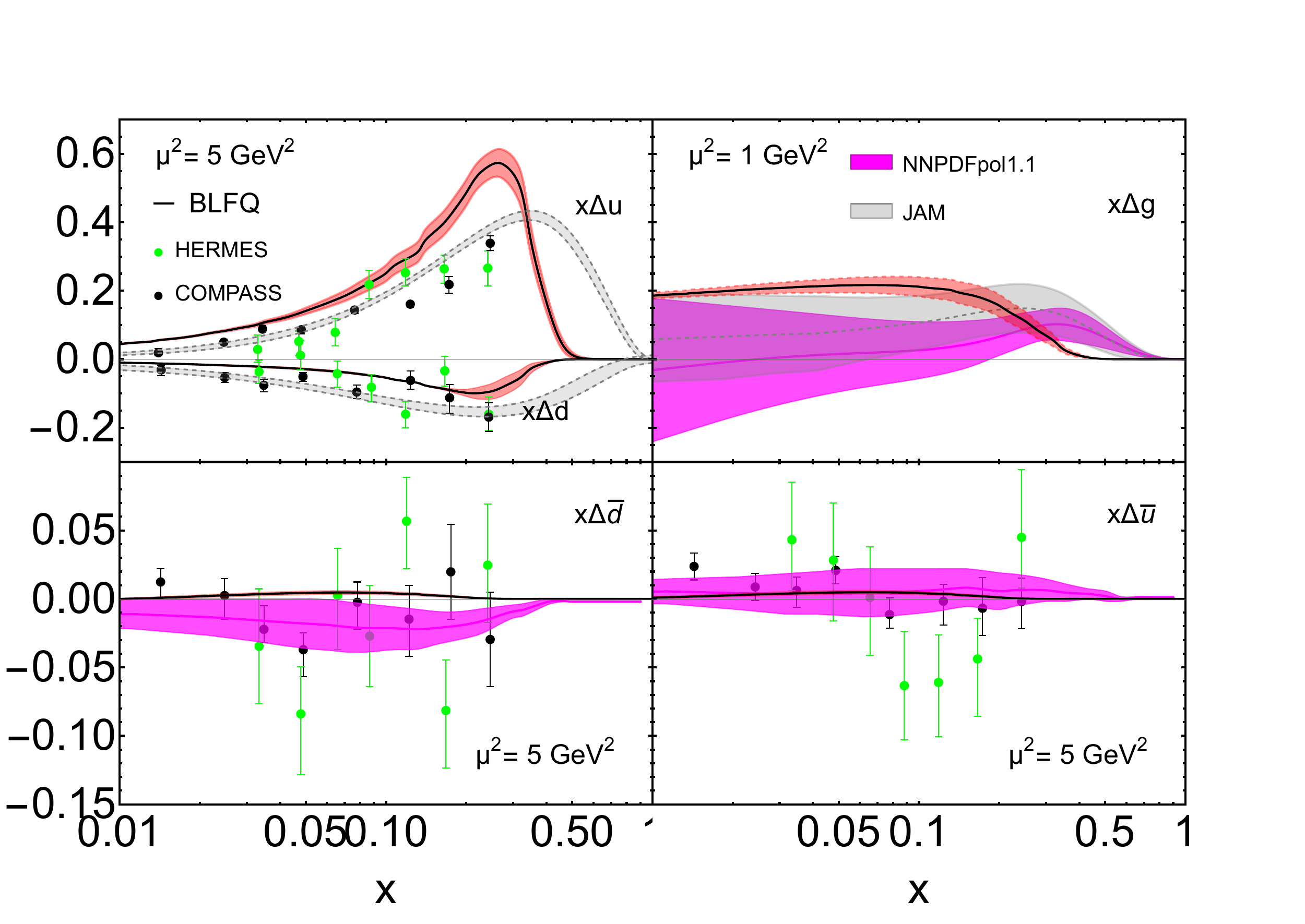}
\caption{Helicity PDFs in the proton computed using LFWFs that include the leading three Fock sectors: \( |qqq\rangle \), \( |qqqg\rangle \), and \( |qqqq\bar{q}\rangle \) as reported in Ref.~\cite{Xu:2024sjt}. Panels show helicity PDFs for valence $u$ and $d$ quarks (upper-left), gluons (upper-right), $\bar{u}$ sea quarks (lower-left), and $\bar{d}$ sea quarks (lower-right). Experimental data are from the COMPASS~\cite{COMPASS:2010hwr} and HERMES~\cite{HERMES:2003gbu,HERMES:2004zsh} Collaborations. The magenta and gray bands correspond to global fits by NNPDFpol1.1~\cite{Nocera:2014gqa} and JAM~\cite{Sato:2016tuz}, respectively.}
\label{fig_hpdf}
\end{center}
\end{figure}

Figure~\ref{fig_hpdf} presents the helicity PDFs of the proton, including valence up and down quarks (upper-left), gluons (upper-right), and sea up and down quarks (lower-left and lower-right, respectively). These distributions are computed using LFWFs derived from a LFQCD Hamiltonian that includes the $|qqq\rangle$, $|qqqg\rangle$, and $|qqqq\bar{q}\rangle$ Fock sectors~\cite{Xu:2024sjt}.

Our predictions for the quark helicity PDFs are in qualitative agreement with experimental data from the COMPASS~\cite{COMPASS:2010hwr} and HERMES~\cite{HERMES:2003gbu,HERMES:2004zsh} collaborations, as well as with global fits from NNPDFpol1.1~\cite{Nocera:2014gqa} and JAM~\cite{Sato:2016tuz}. As in the unpolarized case, the valence quark helicity PDFs appear narrower than those from global analyses for $x > 0.1$. It is also worth noting that the signs of the sea quark helicity distributions remain experimentally undetermined.

Figure~\ref{fig_hpdf} also shows our prediction for the gluon helicity PDF at $\mu^2 = 1~\text{GeV}^2$, compared with global analyses by JAM~\cite{Sato:2016tuz} and NNPDF~\cite{Nocera:2014gqa}. In the small-$x$ region, our result shows reasonable consistency with both fits. At large $x$, the distribution falls off more rapidly than the global results, but it closely tracks the JAM analysis curve up to $x \sim 0.2$. Significant uncertainties remain in the global fits, particularly at small $x$, where even the sign of the gluon helicity PDF is not definitively known~\cite{Zhou:2022wzm}.

The contributions of parton helicities to the proton spin are quantified by the axial charges, defined as the first moments of the helicity PDFs. At the model scale, we obtain axial charges $\Delta \Sigma_u = 0.89 \pm 0.07$ and $\Delta \Sigma_d = -0.22 \pm 0.02$ for the up and down quarks, respectively. While the up-quark contribution agrees well with the global average~\cite{Deur:2018roz}, the down-quark value is somewhat smaller in magnitude than in most global fits. Consequently, the total quark helicity contribution to the proton spin, $\frac{1}{2} \Delta \Sigma = 0.33 \pm 0.04$, is slightly higher than the value suggested by global analyses.

Our results also indicate a significant gluon contribution to the proton spin, with $\Delta G = 0.29 \pm 0.03$ in the range $x_g \in [0.05, 0.2]$ at $\mu^2 = 10$ GeV$^2$. This is in good agreement with the NNPDF global analysis result $\Delta G = 0.23(6)$\cite{Nocera:2014gqa}, as well as with lattice QCD predictions, such as $\Delta G = 0.251(47)(16)$ obtained at the physical pion mass~\cite{Yang:2016plb}.

A similar analysis using only two Fock sectors, $|qqq\rangle$ and $|qqqg\rangle$~\cite{Xu:2022yxb}, yields a total quark helicity contribution of $\frac{1}{2} \Delta \Sigma = 0.359 \pm 0.002$ at the model scale. This includes a dominant up-quark contribution, $\frac{1}{2} \Delta \Sigma_u = 0.438 \pm 0.004$, and a smaller down-quark contribution, $\frac{1}{2} \Delta \Sigma_d = -0.080 \pm 0.002$. While the total $\Delta \Sigma$ and $\Delta \Sigma_u$ agree well with the global averages reported in Ref.~\cite{Deur:2018roz}, the down-quark contribution is noticeably lower than most global extractions. The corresponding gluon helicity contribution in this two-sector scenario is also sizable, with $\Delta G = 0.131 \pm 0.003$.

\begin{figure}
\begin{center}
\includegraphics[width=0.5\linewidth]{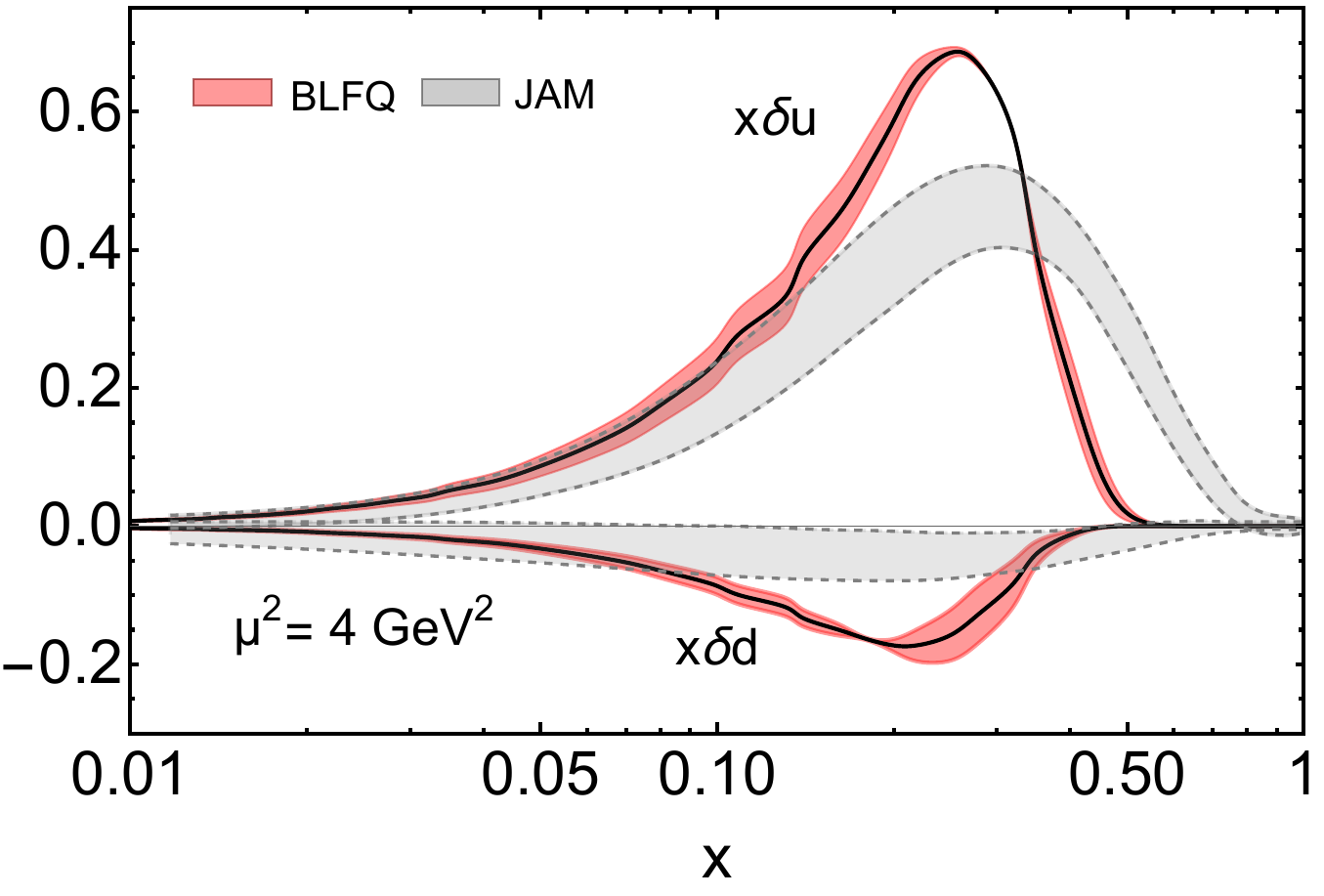}
\caption{Proton transversity PDF computed using LFWFs incorporating the leading three Fock sectors: \( |qqq\rangle \), \( |qqqg\rangle \), and \( |qqqq\bar{q}\rangle \)~\cite{Xu:2024sjt}. BLFQ results are shown as black lines with red bands indicating uncertainties from model parameters, compared to the recent global analysis by JAM~\cite{Cocuzza:2023oam} (gray band).}
\label{fig_h1}
\end{center}
\end{figure}

\subsubsection{Transversity PDFs}
The transversity PDF encodes the distribution of transversely polarized quarks inside a transversely polarized nucleon. In Fig.~\ref{fig_h1}, we present our model predictions for the up and down quark transversity PDFs at $\mu^2 = 4~\text{GeV}^2$, calculated from a  LFQCD Hamiltonian incorporating the $|qqq\rangle$, $|qqqg\rangle$, and $|qqqq\bar{q}\rangle$ Fock sectors, without invoking an explicit confining potential~\cite{Xu:2024sjt}. These results are compared with the recent JAM global analysis~\cite{Cocuzza:2023oam}. Our predictions show good agreement with the JAM fit in the small-$x$ region ($x < 0.1$), while at large $x$, we observe a distinct peak for both up and down quarks—a characteristic also seen in our unpolarized and helicity PDFs.

Notably, the transversity and helicity PDFs in our approach display a near-symmetric pattern for both up and down quarks. This symmetry contrasts with our earlier findings~\cite{Mondal:2019jdg}, where the inclusion of an effective confining potential led to significant flavor asymmetry. We anticipate that including higher Fock components and additional QCD interactions could reintroduce such asymmetry by enriching the spin-flavor dynamics of the nucleon. It is also important to note that, unlike quark transversity, a corresponding gluon transversity PDF is not defined due to the spin-1 nature of the gluon.

The tensor charges, defined as the first moments of the transversity PDFs, provide key information on quark transverse polarization and are of phenomenological importance. At $\mu^2 = 4~\text{GeV}^2$, we obtain $\delta u = 0.81 \pm 0.08$ for the up quark, which moderately exceeds the JAM result $\delta u = 0.71(2)$~\cite{Cocuzza:2023oam} and is consistent with the lattice QCD value $\delta u = 0.784(28)$~\cite{Gupta:2018lvp}. For the down quark, our result $\delta d = -0.22 \pm 0.01$ shows excellent agreement with both the JAM extraction $\delta d = -0.200(6)$ and the lattice prediction $\delta d = -0.204(11)$. These findings suggest that our framework captures the essential nonperturbative features of transversity, and further improvements are expected by systematically including additional Fock sectors and QCD interactions in the Hamiltonian formalism.

\subsection{Generalized Parton Distribution Functions}
GPDs~\cite{Diehl:2003ny} depend on three key kinematic variables: the longitudinal momentum fraction \( x \) carried by the parton, the skewness parameter \( \xi \) that quantifies the longitudinal momentum transfer between the initial and final nucleon, and the invariant momentum transfer squared \( t \). While GPDs are not themselves probability densities, they acquire a probabilistic interpretation in impact-parameter space when evaluated at zero skewness (\( \xi = 0 \)) via a two-dimensional Fourier transform~\cite{Burkardt:2002hr}. Formally, GPDs are defined through off-forward matrix elements of bilocal light-cone operators between nucleon states. The quark GPDs arise from matrix elements involving quark fields, whereas gluon GPDs are constructed from analogous matrix elements of the gluon field strength tensor~\cite{Meissner:2007rx}:
\begin{align}
F^{q}_{\Lambda,\Lambda'} &=\int\frac{dz^-}{4\pi} e^{ix\bar{P}^+z^-}\langle P^\prime,\Lambda^\prime|\bar{\psi}_{q}\big(-\frac{1}{2}z\big)\,\Gamma\,\psi_{q}\big(\frac{1}{2}z\big)|P, \Lambda\rangle \Big{|}_{\substack{z^+=\vec{z}_{\perp}=0}}\nonumber\\
&=\frac{1}{2\bar{P}^+} \bar{u}(P^{\prime},\Lambda^{\prime})\Big[H^q\gamma^+ + E^q \frac{i\sigma^{+\alpha}q_\alpha}{2M}\Big]u(P,\Lambda),\\
F^{g}_{\Lambda,\Lambda'} &= \frac{1}{x\bar{P}^+}\int \frac{dz^-}{2\pi} e^{ix\bar{P}^+z^-} 
\langle P^\prime,\Lambda^\prime|G^{+\mu}\big(-\frac{1}{2}z\big)G^{+}_{\mu}\big(-\frac{1}{2}z\big)|P, \Lambda\rangle \Big{|}_{\substack{z^+=\vec{z}_{\perp}=0}}  \nonumber\\
&=  \frac{1}{2\bar{P}^+} \bar{u}(P',\Lambda')\Big[ H^g \gamma^+ +  E^g  \frac{i\sigma^{+\alpha} q_\alpha}{2M}\Big] u(P,\Lambda), 
\end{align}
where the kinematical variables in the symmetric frame are
$
\bar{P}^\mu=(P^\mu + P^{\prime \mu})/2;~ q^\mu=P^{\prime \mu}-P^\mu;~ \xi=-q^+/2\bar{P}^+;\nonumber
$
and $t=q^2$. For $\xi=0$, $t=-\bm{q}_{\perp}^2=-Q^2$.

The unpolarized GPDs at zero skewness ($\xi = 0$) can be represented using LFWFs via the overlap formalism as
\begin{align}\label{eq_H}
H^{q/g}(x,0,t) &= \frac{1}{2} \int_{N} \Psi^{N,\,\Lambda\,*}_{\{x_i^\prime,\bm{p}_{i\perp}^{\,\prime},\lambda_i\}} \, \Psi^{N,\,\Lambda}_{\{x_i,\bm{p}_{i\perp},\lambda_i\}} \delta (x-x_1), \\
E^{q/g}(x,0,t) &= -\frac{M}{(q^1-iq^2)} \int_{N} \Psi^{N,\,\Lambda\,*}_{\{x_i^\prime,\bm{p}_{i\perp}^{\,\prime},\lambda_i\}} \, \Psi^{N,\,-\Lambda}_{\{x_i,\bm{p}_{i\perp},\lambda_i\}} \delta (x-x_1).
\end{align}

\begin{figure}
\begin{center}
\includegraphics[width=0.3\linewidth]{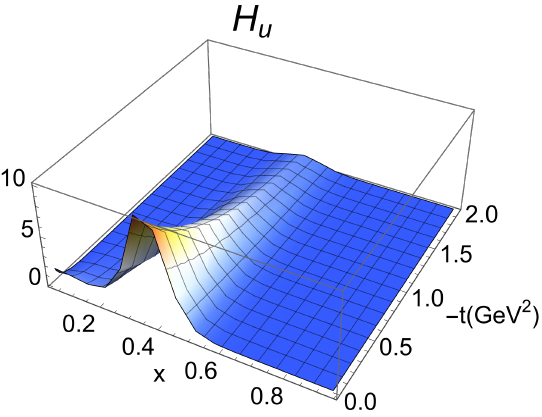}
\includegraphics[width=0.3\linewidth]{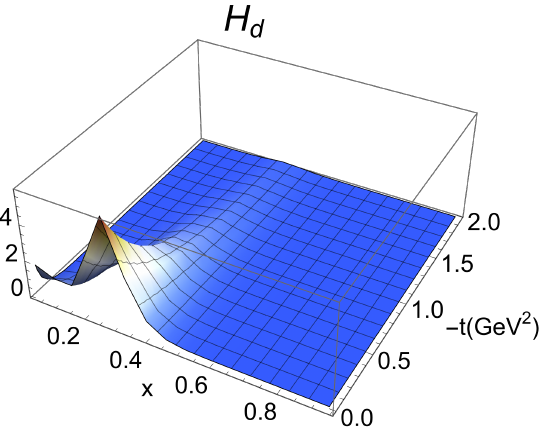}
\includegraphics[width=0.31\linewidth]{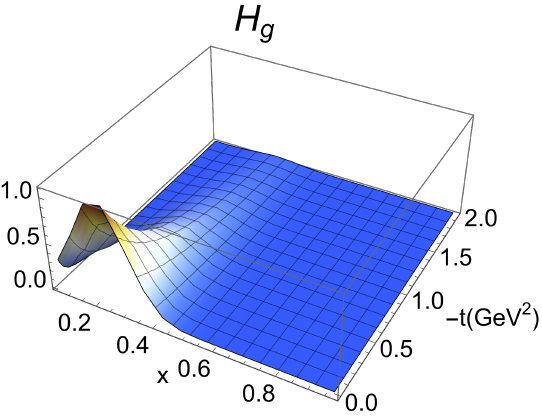}\\
\includegraphics[width=0.3\linewidth]{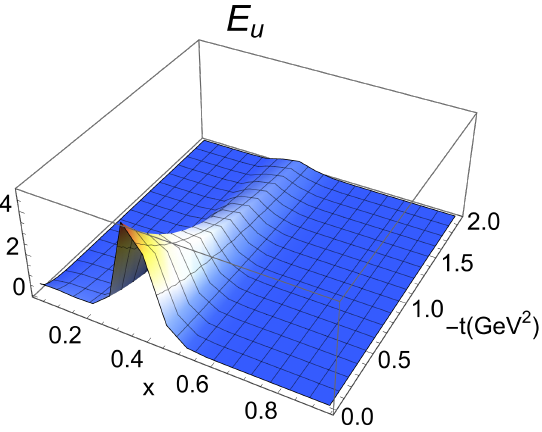}
\includegraphics[width=0.3\linewidth]{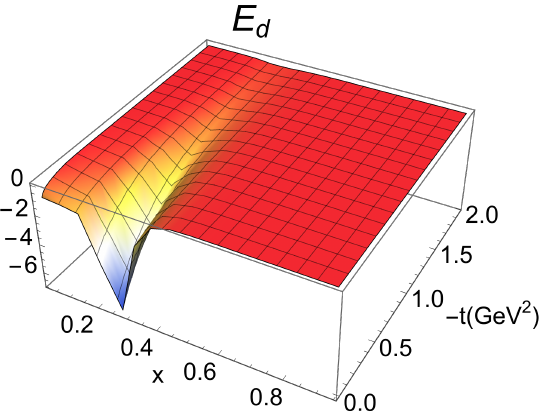}
\includegraphics[width=0.31\linewidth]{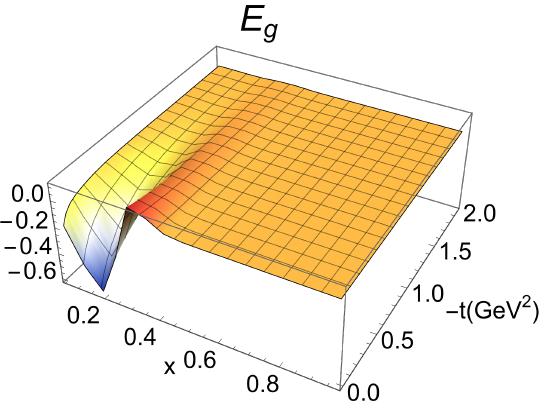}
\caption{Unpolarized GPDs \(H(x,0,t)\) and \(E(x,0,t)\) for the \(u\) quark (left panel), \(d\) quark (middle panel), and gluon (right panel), shown as functions of \(x\) and \(-t\). These results are computed using LFWFs that include the leading three Fock sectors: \( |qqq\rangle \), \( |qqqg\rangle \), and \( |qqqq\bar{q}\rangle \)~\cite{Xu:2024sjt}.}
\label{GPDs}
\end{center}
\end{figure}

The quark and gluon GPDs of the proton have been extensively studied within the BLFQ framework~\cite{Xu:2021wwj,Kaur:2023lun,Liu:2022fvl,Zhang:2023xfe,Liu:2024umn,Lin:2024ijo,Lin:2023ezw,Zhang:2025nll}. Earlier investigations were restricted to the valence Fock sector composed of three constituent quarks~\cite{Xu:2021wwj}, or extended to include a single additional Fock component containing a dynamical gluon~\cite{Xu:2022yxb}. In this review, we focus on the most recent advancements that incorporate a broader Fock space expansion. In particular, we present GPDs derived from LFWFs constructed using an extended basis that includes the three-quark, three-quark–gluon, and three-quark–quark–antiquark sectors~\cite{Xu:2024sjt,Mondal:2025bjn}, providing a more comprehensive representation of the proton’s nonperturbative structure.

In Fig.~\ref{GPDs}, we show the unpolarized GPDs \( H \) and \( E \) for individual quark flavors and for the gluon at the model scale, plotted as functions of the light-cone momentum fraction \( x \) and the squared momentum transfer \( -t \). These distributions exhibit a peak when the proton transfers no transverse momentum and the struck parton carries less than half of the proton’s longitudinal momentum. As the transverse momentum transfer increases, the peak shifts to larger \( x \), and the overall magnitude diminishes. At large \( x \), both \( H \) and \( E \) fall off and become nearly independent of \( t \), with \( E \) decreasing more rapidly than \( H \). Notably, \( E^d \) is negative, reflecting the negative anomalous magnetic moment of the down quark.

These qualitative trends are consistent with results from a range of QCD-inspired models~\cite{Liu:2022fvl,Mondal:2015uha,Pasquini:2006dv,Chakrabarti:2013gra,deTeramond:2018ecg,Lin:2023ezw,Maji:2017ill,Kaur:2023lun,Liu:2024umn,Chakrabarti:2023djs,Zhang:2025nll} as well as lattice QCD calculations~\cite{Alexandrou:2020zbe,Lin:2020rxa,Bhattacharya:2022aob}, suggesting a degree of model independence. However, the use of relatively large constituent quark masses in our model, together with the omission of higher Fock sectors and a longitudinal confining potential, limits the description of longitudinal excitations. Consequently, our GPDs tend to be narrower than those from lattice QCD and other approaches, particularly in the region \( x > 0.1 \).

\subsubsection{Orbital Angular Momentum}

The contributions of partonic OAM to the proton spin in the light-cone gauge, $A^+=0$, can be evaluated within the GPD framework via the sum rule~\cite{Ji:1996ek}:
\begin{equation}
L_z = \int \mathrm{d}x \left\{ \frac{1}{2} x \left[ H(x,0,0) + E(x,0,0) \right] - \tilde{H}(x,0,0) \right\}.
\end{equation}
Using this expression and the Fock space consisting of \( |qqq\rangle \), \( |qqqg\rangle \), and \( |qqqq\bar{q}\rangle \)~\cite{Xu:2024sjt}, we obtain \( L_z^u = 0.036 \pm 0.002 \) for up quarks, \( L_z^d = 0.058 \pm 0.001 \) for down quarks, and \( L_z^g = -0.037 \pm 0.004 \) for gluons.

Although a direct experimental determination of OAM remains elusive, several promising approaches have been proposed. Gluon OAM may be probed through double spin asymmetries in diffractive dijet production~\cite{Bhattacharya:2022vvo}, while quark OAM could potentially be accessed via exclusive double Drell-Yan processes~\cite{Bhattacharya:2017bvs}.

\begin{figure}[htbp]
    \centering
  \includegraphics[scale=0.44]{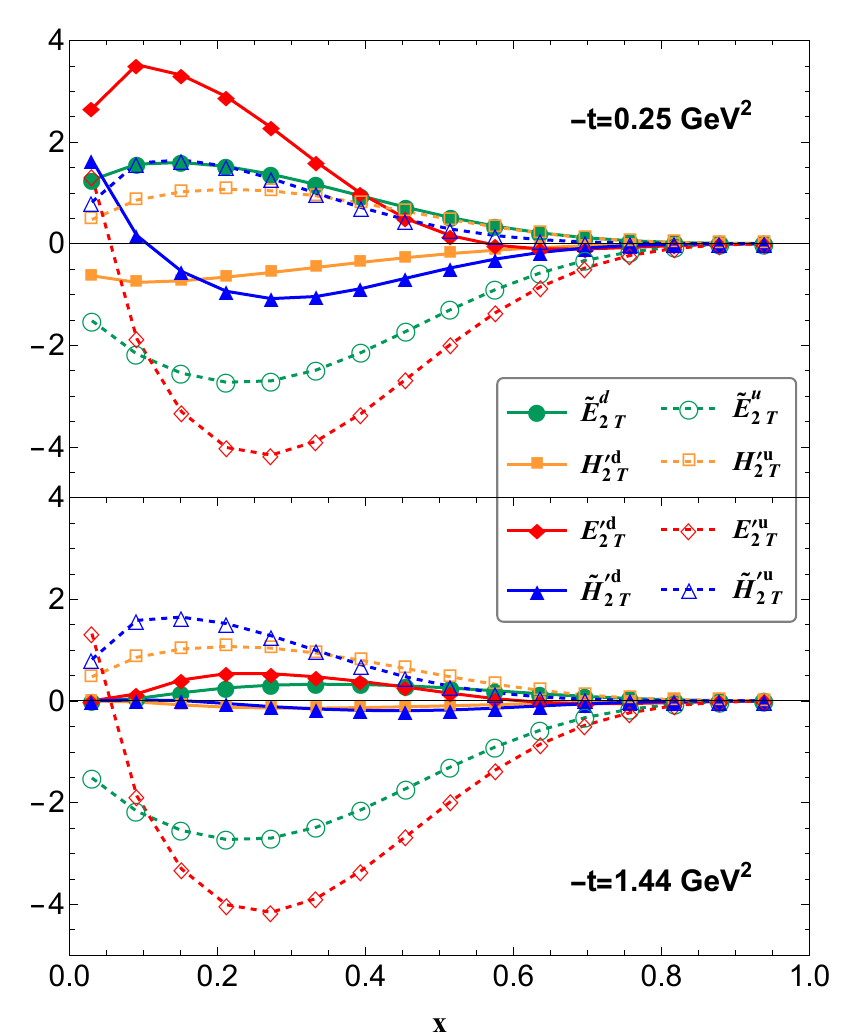} 
  \includegraphics[scale=0.44]{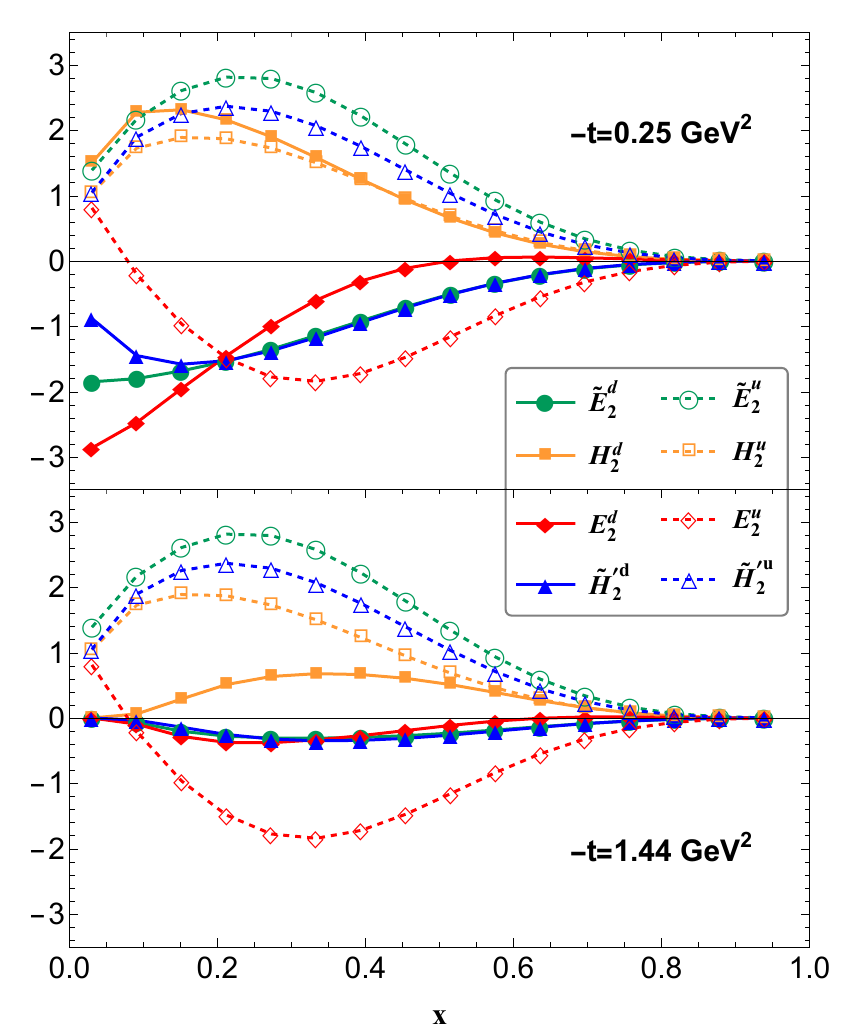}
    \caption{BLFQ calculations for the proton twist-3 GPDs from Ref.~\cite{Zhang:2023xfe}. The upper and lower rows show the twist-3 GPDs at fixed values of $-t = 0.25\ \mathrm{GeV}^2$ and $-t = 1.44\ \mathrm{GeV}^2$, respectively, at the quark level. The corresponding flavor-level distributions are obtained using $X_{\text{flavor}}^u = 2 X_{\text{quark}}^u$ and $X_{\text{flavor}}^d = X_{\text{quark}}^d$. In each panel, the same color is used for a given GPD across flavors, with solid lines and filled markers representing down-quark contributions, and dashed lines with open markers indicating up-quark contributions.
} 
    \label{first2dx}
\end{figure}

\begin{figure}[htbp]
    \centering
    \includegraphics[scale=0.5]{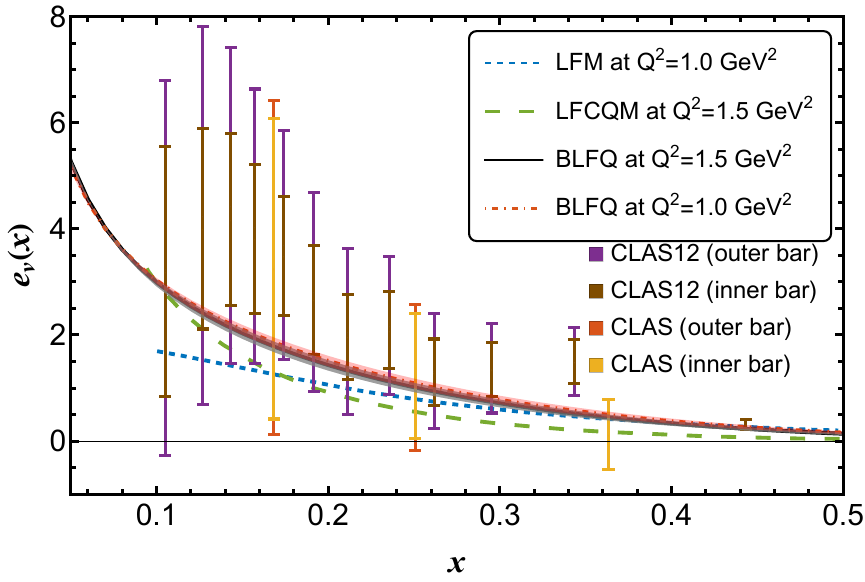}
    \caption{BLFQ calculations for the proton flavor combination 
$e_v = \frac{4}{9} (e_u - e_{\bar{u}}) - \frac{1}{9} (e_d - e_{\bar{d}})$ 
at $Q^{2} = 1.0~\mathrm{GeV}^{2}$ (red line) and $Q^{2} = 1.5~\mathrm{GeV}^{2}$ (black line)~\cite{Zhang:2023xfe}, shown in comparison with other theoretical and experimental results. The blue and green lines represent model calculations from Refs.~\cite{Pasquini:2018oyz} and~\cite{Lorce:2014hxa}, respectively. The purple and brown error bars correspond to extractions from CLAS data, while the red and yellow error bars are obtained from CLAS12 measurements. The inner error bars indicate the contribution from the zeroth-order approximation, while the outer bars include twist-3 corrections in the fragmentation sector, as detailed in Ref.~\cite{Courtoy:2022kca}.
}
    \label{evloved}
\end{figure}

\subsubsection{Subleading-Twist GPDs}
Beyond the leading-twist GPDs, we have computed all well-defined twist-3 GPDs—both chiral-even and chiral-odd—in the zero-skewness limit of the proton, within the BLFQ framework~\cite{Zhang:2023xfe}. These twist-3 GPDs are obtained using overlap representations constructed from LFWFs, with calculations performed in a truncated Fock space (retaining the leading Fock component) and regulated by numerical cutoffs. Key features of the GPDs are highlighted through 2D forms in Fig.~\ref{first2dx}, where $-t$ is fixed at $0.25 \ \mathrm{GeV}^2$ and $1.44 \ \mathrm{GeV}^2$. One finds that the peaks are moving towards higher $x$ as $-t$ increases.

We have also evaluated the corresponding twist-3 PDFs, including the spin-independent distribution $e(x)$, which we have evolved to a higher scale for comparison with other theoretical models and available experimental data~\cite{Courtoy:2022kca,Pasquini:2018oyz,Lorce:2014hxa}. Our results for $e(x)$ shown in Fig.~\ref{evloved} exhibit general consistency among selected models~\cite{Pasquini:2018oyz,Lorce:2014hxa} and reasonable agreement with the CLAS experimental extractions~\cite{Courtoy:2022kca}, while acknowledging the substantial uncertainties in the data.

In future work, we aim to extend these studies by incorporating higher Fock sectors into the calculation of twist-3 GPDs and exploring their connection to the proton’s OAM. The investigation of nonzero-skewness cases is also of significant interest, as such observables are related to twist-3 contributions to DVCS cross sections~\cite{Guo:2022cgq} and may be accessed experimentally at facilities such as the EIC and EicC~\cite{Chavez:2021koz}. These future developments at twist-3 and finite skewness will provide deeper insights into the three-dimensional structure of the proton and contribute to resolving the proton spin puzzle.

\subsection{Transverse-momentum-dependent parton distributions}
The TMDs~\cite{Boussarie:2023izj} extend the concept of collinear PDFs by providing three-dimensional information about the internal momentum structure of hadrons. In addition to encoding the partons’ longitudinal momentum fractions, TMDs capture correlations between the spin of the target hadron and the transverse momenta of its constituent partons. At leading twist, there are eight TMDs for the nucleon. Among them, \( f_1(x, p_\perp^2) \), \( g_{1L}(x, p_\perp^2) \), and \( h_1(x, p_\perp^2) \) are direct generalizations of the corresponding leading-twist collinear PDFs. The remaining TMDs, however, do not have simple collinear limits and reflect genuinely transverse-momentum--dependent dynamics.

TMDs provide a framework to describe a wide range of phenomena within QCD through established factorization theorems~\cite{Rogers2016,Collins2011,Collins1985,Collins1981,Collins1982,Aybat2011,Aybat2012}. For example, the Collins asymmetry observed in SIDIS can be interpreted using the transversity TMD, \( h_1(x, p_\perp^2) \)~\cite{Anselmino2015,Kang2016,DAlesio2020,Cammarota2020}. The double spin asymmetry \( A_{LT} \) in SIDIS is described by the worm-gear TMD, \( g_{1T}(x, p_\perp^2) \)~\cite{Bhattacharya2021}, while the single spin asymmetry \( A_{UT}^{\sin(3\phi_h - \phi_S)} \) is related to the pretzelosity TMD, \( h_{1T}^{\perp}(x, p_\perp^2) \)~\cite{Lefky2015a}.

We compute the quark TMD PDFs of the proton from its LFWFs within the framework of  BLFQ~\cite{Hu:2022ctr}. These LFWFs are obtained as eigenvectors of an effective light-front Hamiltonian truncated to the leading Fock sector, incorporating a three-dimensional confining potential and a one-gluon exchange interaction with fixed coupling.

\begin{figure}
\centering
    \includegraphics[width=0.6\textwidth]{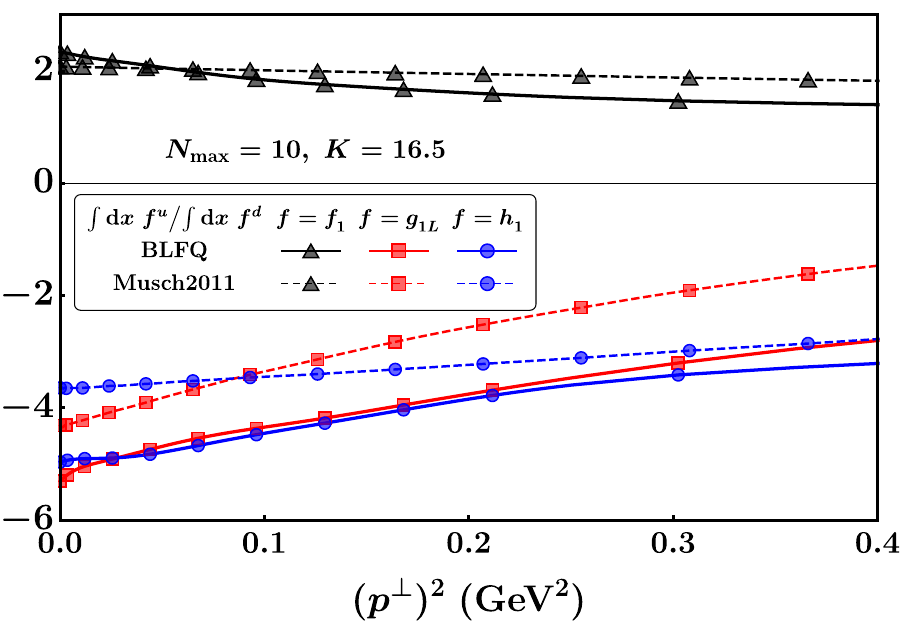}
    \caption{\label{fig:xintegration_falourratio_comparison_f1g1Lh1} Comparison of the integrated flavor ratios \( \left. \int \mathrm{d}x\, f^u \middle/ \int \mathrm{d}x\, f^d \right. \) for the TMDs \( f_1 \), \( g_{1L} \), and \( h_1 \), as computed within the BLFQ framework and from lattice QCD simulations. The lattice results correspond to the central values obtained by parameterizing \( \widetilde{A}_2 \), \( \widetilde{A}_6 \), and \( \widetilde{A}_{9m} \)~\cite{Musch2011}. Lines with different markers (colors) represent the flavor ratios for each distribution, as indicated in the legend. Solid lines denote the BLFQ results, while dashed lines correspond to the lattice QCD predictions.
}
\end{figure}

In our study, the gauge link is set to unity, restricting the analysis to the T-even sector and resulting in six nonzero leading-twist TMDs out of the full set of eight. We compare our results within the same BLFQ framework and with lattice QCD simulations, as shown in Fig.~\ref{fig:xintegration_falourratio_comparison_f1g1Lh1}, and observe good agreement between them. The validity of the universal Soffer-type inequalities~\cite{Bacchetta2000}, along with the absence of previously identified model-dependent relations~\cite{Bacchetta2007,Goeke2003,Mulders1995}, suggests that the BLFQ framework captures essential features of nonperturbative QCD dynamics. Including additional Fock sectors would introduce more independent helicity amplitudes, enabling the emergence of further independent TMDs, including those of higher twist and T-odd nature. Such extensions are expected to move us closer to a complete nonperturbative description of QCD.

As demonstrated in Ref.~\cite{Hu:2022ctr}, our calculations do not support the commonly assumed \( x\text{--}p_\perp \) factorization used in several phenomenological studies~\cite{Anselmino2014,Bhattacharya2021,Anselmino2015,DAlesio2020,Cammarota2020,Lefky2015a}. In particular, the nontrivial \( x \)-dependence of the transverse momentum moments \( \langle (p_\perp)^2 \rangle_f^q \) rules out the factorized ansatz of the form
\begin{align}
    f^q(x, p_\perp^2) = f^q(x) \frac{\hat{f}^q(p_\perp^2)}{\mathcal{N}} \, ,
\end{align}
where \( \mathcal{N} = \int \mathrm{d}^2 p^\perp\, \hat{f}^q(p_\perp^2) \). Furthermore, comparison between the BLFQ results and Gaussian-type parameterizations indicates that the Gaussian ansatz is only effective in describing the TMDs in the small-\( p_\perp^2 \) region~\cite{Hu:2022ctr}.

We have also calculated the T-even gluon TMDs at leading twist within the BLFQ framework~\cite{Yu:2024mxo}. The LFWFs of the proton used in this analysis are obtained from a light-front quantized QCD-inspired Hamiltonian. The model includes both the valence Fock sector with three constituent quarks and an additional sector comprising three quarks and a dynamical gluon, with a three-dimensional confining interaction in the valence Fock component~\cite{Xu:2022yxb}. Our results are consistent with available theoretical predictions and experimental extractions. However, unlike the findings reported in Refs.~\cite{Bacchetta2020,Chakrabarti:2023djs,Lyubovitskij:2020xqj}, the gluon TMDs obtained in BLFQ are dominated by S-wave components, resulting in monotonic behavior in the transverse momentum direction. 

Moreover, due to the asymptotic behavior in transverse momentum of all four T-even gluon TMDs computed in our approach, the associated polarized gluon densities do not exhibit significant asymmetries. Whether the inclusion of higher Fock sectors in BLFQ introduces sizable asymmetries in gluon polarized densities remains an open question, which we aim to explore in future work.

Our calculations of gluon TMDs~\cite{Yu:2024mxo} do not support the model relation proposed in Ref.~\cite{Lyubovitskij:2020xqj}, which connects the square of \( f_1^g(x, p_\perp^2) \) to the sum of the squares of the other three T-even gluon TMDs. In contrast, certain constraints on gluon TMDs can be derived directly from QCD principles. We examine two such constraints: (i) the Mulders–Rodrigues inequalities~\cite{Cotogno:2017puy,Mulders:2000sh}, and (ii) the asymptotic behaviors of TMD moments based on Refs.~\cite{Boer:2016xqr,Brodsky:1989db,Brodsky:1994kg}. The T-even gluon TMDs calculated within the BLFQ framework satisfy all of these QCD-based constraints, reinforcing the credibility of our results and supporting future investigations of gluon TMD-related observables in experiments.


\subsubsection{Subleading-Twist TMDs}
Although the twist-3 contribution to the cross section of high-energy scattering processes is suppressed by a factor of \(1/Q\), this suppression does not imply that twist-3 distributions themselves are small. In fact, as indicated by the equation of motion (EOM) relations~\cite{Efremov:2002qh,Bacchetta:2006tn,Lorce:2014hxa}, twist-3 distributions can be comparable in magnitude to their twist-2 counterparts~\cite{Lorce:2014hxa, Pasquini:2018oyz, Bhattacharya:2020cen, Bhattacharya:2021moj, Zhu:2023lst}. These distributions become experimentally accessible in kinematic regimes where \(Q\) is moderate, particularly in the context of upcoming facilities such as the EicC~\cite{Anderle:2021wcy,Anderle:2021dpv,Zeng:2022lbo,Zeng:2023nnb} and the EIC~\cite{AbdulKhalek:2021gbh, Burkert:2022hjz}. However, due to the inherently nonperturbative nature of QCD, as well as the incomplete understanding of color confinement and chiral symmetry breaking, determining twist-3 distributions from first principles remains a formidable challenge.

We performed a systematic calculation of the subleading structure functions of the proton, with particular emphasis on the interference terms corresponding to the genuine twist-3 distribution functions~\cite{Zhu:2024awq}. In our approach, we adopted two simplifying approximations: the omission of zero modes and the replacement of the gauge link by the identity matrix. These approximations eliminate the T-odd TMDs and the singular terms arising in the EOM relations.

We first computed the twist-2 T-even TMDs and the genuine twist-3 T-even TMDs using the LFWFs of the proton obtained within the BLFQ framework. The genuine twist-3 distributions arise from the overlap of LFWFs associated with the \( |qqq\rangle \) and \( |qqqg\rangle \) Fock sectors.

Using the EOM relations, we then constructed the full twist-3 T-even TMDs. Finally, by integrating the relevant TMDs over transverse momentum \( \boldsymbol{p}_\perp \), we obtained the corresponding twist-3 PDFs.

%
At the model scale, our results indicate that the genuine twist-3 distributions are generally smaller in magnitude compared to the total twist-3 distributions. Notably, enhanced contributions appear in the small-\(x\) region suggesting the presence of strong interference effects at low longitudinal momentum fractions~\cite{Zhu:2024awq}. 

The total twist-3 distributions themselves are sizable at small \(x\), indicating that experiments at the EicC, operating in the small-\(x\) and low-\(Q^2\) regime, may be sensitive to higher-twist effects such as the azimuthal asymmetry \( A_{LT}^{\cos\phi_S} \)~\cite{Bacchetta:2006tn,Wang:2016dti}. 

An intriguing outcome of our analysis is the apparent violation of the Burkhardt–Cottingham (BC) sum rules~\cite{Burkhardt:1970ti,Tangerman:1994bb,Burkardt:1995ts}, underscoring the potential importance of zero-mode contributions in the calculation of twist-3 distributions.

Moreover, the fulfillment of positivity bounds for the twist-2 distributions, along with the observed independence among all computed distributions, further supports the reliability of the BLFQ framework in capturing essential nonperturbative QCD dynamics.

We anticipate that these results will advance our understanding of the proton’s partonic structure, particularly the internal correlations and interference effects among multiple partons. In addition, they provide valuable nonperturbative input for future experimental efforts aimed at probing higher-twist phenomena at the EIC and EicC.

\section{Nucleon Structure Using Full BLFQ}
\label{sec:fullBLFQ}

Recent applications of BLFQ to nucleon structure have employed explicit truncations of the Fock space. The inclusion of a dynamical gluon, beyond the valence three-quark sector, has already led to notable improvements, enhancing agreement with experimental observables. Most notably, BLFQ has recently reached a major milestone by computing nucleon LFWFs as eigenstates of the QCD Hamiltonian without introducing an explicit confining potential. These calculations incorporate Fock sectors up to the five-parton state \(|qqqq\bar{q}\rangle\), enabling first-principles predictions of quark and gluon distributions, helicity and transversity PDFs, and spin-dependent observables. The results exhibit qualitative consistency with both experimental data and global analyses.

A key future direction is the realization of Full BLFQ~\cite{Vary:2009gt}, in which the LFQCD Hamiltonian is solved nonperturbatively using basis-space regulators alone, without imposing additional truncation in Fock space. Removing such truncations represents a critical step toward establishing BLFQ as a truly \emph{ab initio} approach to solving QCD. Initial demonstrations of Full BLFQ have been carried out in simpler settings, such as scalar field theories in \((1+1)\) dimensions~\cite{Vary:2021cbh}, where solutions have been obtained using only a basis space regulator without an additional Fock space truncation.

The Full BLFQ framework treats the nucleon as a quantum many-body system with a variable number of partons. The preferred basis consists of single-particle HO states combined with discretized longitudinal momentum modes, regulated by the \(N_{\max}\)–\(K\) scheme:
\begin{equation}
\begin{cases}
\sum_i \left(2 n_i + |m_i| + 1 \right) \leq N_{\max}, \\
\sum_i p_i^+ = \frac{2\pi K}{L}.
\end{cases}
\end{equation}
This choice preserves all kinematical symmetries of the LFQCD Hamiltonian, including exact factorization of the CM motion within the many-body Hilbert space. The basis implements soft IR and UV cutoffs in a transverse momentum as well as longitudinal momentum discretization characterized by:
\begin{equation}
\begin{aligned}
& \frac{b^2}{N_{\max} - 1} \lesssim \sum_i \frac{k_{i \perp}^2}{x_i} \lesssim b^2 (N_{\max} - 1), \\
& \Delta x \gtrsim \frac{1}{K},
\end{aligned}
\end{equation}
where \(b = \sqrt{P^+ \Omega}\), with \(P^+\) the total longitudinal momentum of the bound state and \(\Omega\) the transverse HO scale parameter. It is important to note that, when zero modes are neglected as is standard the \(N_{\max}\)--\(K\) regularization renders the parton number finite, eliminating the need for invoking a Fock sector truncation.

A central challenge in full BLFQ is the exponential growth of the Hilbert space dimension, \(\dim \mathcal{H} = N^{dN}\), where \(N = \max\{N_{\max}, K\}\) and \(d\) is the single-particle state dimension. This rapid scaling is a common feature of strongly coupled, nonperturbative quantum many-body problems. Nevertheless, ongoing advances in high-performance computing, including exascale systems capable of \(10^{18}\) floating-point operations per second, offer hope for practical calculations. Moreover, emerging quantum computing technologies promise to surpass classical computational capabilities, providing potentially transformative approaches to tackling the complexity inherent in full BLFQ~\cite{Kreshchuk:2020dla,Qian:2021jxp,Kreshchuk:2020kcz}.

\subsection{Emergent Hadronic Mass in Full BLFQ}
A central challenge for advancing the BLFQ framework toward a full QCD realization is the incorporation of emergent hadronic mass (EHM)—the fundamental QCD mechanism by which most of the nucleon mass is generated dynamically from gluon and quark interactions rather than arising from the small current masses of the $u$ and $d$ quarks~\cite{Binosi:2025kpz,Achenbach:2025kfx}. 
The studies reviewed in this article have so far employed Fock-space truncations and effective constituent-quark masses, as is typical of phenomenological light-front models. These approximations enable quantitative descriptions of spectroscopy and partonic observables, yet they do not reproduce the full dynamical origin of hadronic mass or the underlying mechanisms of confinement and chiral symmetry breaking that are intrinsic to QCD. 

Realizing EHM within Full BLFQ will require several conceptual and computational advances. The inclusion of additional dynamical gluon and sea-quark degrees of freedom is essential to recover the complete nonperturbative dynamics of QCD. Furthermore, renormalization procedures must be developed that are consistent with the QCD $\beta$ function and that preserve the correct running of the strong coupling across different scales. In addition, the  light-front Hamiltonian must incorporate the infrared structure associated with confinement and dynamical chiral symmetry breaking, possibly through connections to Dyson--Schwinger, Bethe--Salpeter, or light-front holographic approaches. 

Achieving a consistent description of EHM within the BLFQ framework would constitute a major milestone in the pursuit of a first-principles, Hamiltonian-based understanding of nucleon structure. Such progress would bridge the gap between the fundamental Lagrangian of QCD and the emergent mass and structure of hadrons, establishing Full BLFQ as a comprehensive nonperturbative tool for strong-interaction physics.

\section{Conclusion and Outlook}

In this review, we have summarized recent progress in the application of BLFQ to the study of nucleon structure. Through a series of systematic developments, the fully relativistic and non-perturbative  BLFQ framework has evolved from early studies with minimal Fock content to incorporate essential features of LFQCD dynamics. Notably, the inclusion of the \( |qqq\rangle \), \( |qqqg\rangle \), and \( |qqqq\bar{q}\rangle \) Fock sectors enables an increasingly realistic description of the nucleon's partonic structure. These advancements have yielded predictions for a broad range of observables, including electromagnetic form factors, PDFs, GPDs, and spin-dependent quantities that exhibit encouraging agreement with experimental measurements and global QCD analyses.

In particular, the recent realization of nucleon LFWFs as eigenstates of a QCD-based Hamiltonian without an explicit confining potential marks a key milestone. These wave functions serve as a common foundation for computing diverse observables such as quark and gluon helicity and transversity distributions, tensor charges, and the decomposition of the proton spin. The results obtained demonstrate the power of the BLFQ approach to reveal insights into the interplay of quark and gluon degrees of freedom within the nucleon.

Looking ahead, several promising directions for further development emerge. First, the inclusion of higher Fock sectors especially those involving multi-gluon configurations is essential for a more complete treatment of non-Abelian QCD dynamics and confinement mechanisms. Second, ongoing efforts to refine the Hamiltonian with scale-dependent couplings and improved renormalization will further enhance the precision and predictiveness of the framework.

The long-term goal is to realize Full BLFQ, where the QCD Hamiltonian is solved nonperturbatively using basis-space regulators alone, free from invoking a Fock sector truncation. This would elevate BLFQ to a fully \emph{ab initio} method for solving QCD. Initial success in simpler theories suggests that this goal is within reach, especially with the advent of exascale computing and the emergence of quantum simulation platforms for quantum field theory.


Finally, the LFWFs generated within this framework provide a versatile platform for future studies of three-dimensional partonic structure, encompassing azimuthal spin asymmetries in SIDIS and Drell–Yan processes, photo- and electroproduction of vector quarkonia off the proton, Wigner distributions, and multi-parton correlations. With the upcoming experimental programs such as the EICs, the interplay between theoretical advances and high-precision data will be essential for deepening our understanding of nucleon structure. In this context, the BLFQ approach emerges as a powerful tool, offering a path toward a unified and comprehensive description of hadronic matter in QCD.

\section*{Acknowledgement} 
We thank Satvir Kaur for preparing Fig.~\ref{publications} presented in this article and for many fruitful discussions. CM is supported by new faculty start up funding by the Institute of Modern Physics, Chinese Academy of Sciences, Grant No.~E129952YR0.  CM also thanks the Chinese Academy of Sciences Presidents International Fellowship Initiative for support via Grants No.~2021PM0023. XZ is supported by the National Natural Science Foundation of China under Grant No.~12375143, by new faculty start-up funding by the Institute of Modern Physics, Chinese Academy of Sciences, by the Key Research Program of Frontier Sciences, Chinese Academy of Sciences, Grant No.~ZDBS-LY-7020, by the Natural Science Foundation of Gansu Province, China, Grant No.~20JR10RA067, by the Foundation for Key Talents 
of Gansu Province, by the Central Funds Guiding the Local Science and Technology 
Development of Gansu Province, Grant No.~22ZY1QA006, by the International Partnership 
Program of the Chinese Academy of Sciences, Grant No.~016GJHZ2022103FN, by the Strategic 
Priority Research Program of the Chinese Academy of Sciences, Grant No.~XDB34000000, 
and by National Key R\&D Program of China, Grant No. 2023YFA1606903. This research is also supported by Gansu International Collaboration and Talents Recruitment Base of Particle Physics (2023-2027), by the Senior Scientist Program funded by Gansu Province, Grant No.~25RCKA008. YL is supported by the new faculty startup fund of University of Science and Technology of China, by the NSFC under Grant No. 12375081, by the Chinese Academy of Sciences under Grant
No. YSBR-101. 
This work is also supported by Gansu International Collaboration and Talents Recruitment Base of Particle Physics (2023–2027), by the Senior Scientist Program funded by Gansu Province Grant No. 25RCKA008.

\bibliographystyle{elsarticle-num}
\bibliography{Review_BLFQ_Nucleon_9Nov.bib}

\end{document}